%% file: aa51585-24corr.tex
\documentclass{aa} 

\usepackage{graphicx}
\usepackage[varg]{txfonts}
\usepackage[breaklinks, colorlinks, citecolor=blue, linkcolor=blue]{hyperref}
\usepackage{longtable}
\usepackage{float,lscape} 
\usepackage{caption}
\usepackage{amsmath}
\usepackage{natbib}

\bibpunct{(}{)}{;}{a}{}{,} 
\newcommand{\tgray}[1]{\textcolor[rgb]{0.4,0.4,0.4}{#1}}

\begin{document} 

 \title{CARMENES input catalogue of M dwarfs}
 \subtitle{VIII. Kinematics in the solar neighbourhood}
%
 \titlerunning{Kinematics of M dwarfs in the solar neighbourhood}

\author{
    M.~Cort\'es-Contreras \inst{1,2}
    \and
    J.\,A.~Caballero \inst{2}
    \and
    D.~Montes \inst{1}
    \and
    C.~Cardona-Guill\'en \inst{3,4}
    \and
    V.\,J.\,S.~B\'ejar \inst{3,4}
    \and 
    C.~Cifuentes \inst{2}
    \and
    H.\,M.~Tabernero \inst{1,2}
    \and
    M.\,R.~Zapatero Osorio \inst{2}
    \and
    P.\,J.~Amado \inst{5}
    \and
    S.\,V.~Jeffers \inst{6}
    \and
    M.~Lafarga\inst{7,8}
    \and
    N.~Lodieu \inst{3,4}
    \and        
    A.~Quirrenbach \inst{9}
    \and
    A.~Reiners \inst{10}
    \and
    I.~Ribas \inst{11,12}
    \and
    P.~Sch\"ofer \inst{5}
    \and
    A.~Schweitzer \inst{13}
    \and
    W.~Seifert \inst{9}
} 

  \authorrunning{M.~Cort\'es-Contreras~et~al.}

 \institute{
 Departamento de F\'isica de la Tierra y Astrof\'isica, Facultad de Ciencias F\'isicas, e IPARCOS-UCM (Instituto de Física de Partículas y del Cosmos de la UCM), Universidad Complutense de Madrid, 28040 Madrid, Spain, \email{mcortescontreras@ucm.es}
\and
        Centro de Astrobiolog\'ia (CSIC-INTA), Camino Bajo del Castillo s/n, 28691 Villanueva de la Ca\~nada, Madrid, Spain 
\and
        Instituto de Astrof\'isica de Canarias, V\'ia L\'actea s/n, 38205 La Laguna, Tenerife, Spain
\and Departamento de Astrof\'isica, Universidad de La Laguna, 38206 La Laguna, Tenerife, Spain
\and
        Instituto de Astrof\'isica de Andaluc\'ia (CSIC), Glorieta de la Astronom\'ia s/n, 18008 Granada, Spain
\and
Max-Planck Institute for Solar System Research, Justus-von-Liebig-Weg 3, 37077 G\"ottingen, Germany
\and
Department of Physics, University of Warwick, Gibbet Hill Road, Coventry CV4 7AL, United Kingdom
\and
Centre for Exoplanets and Habitability, University of Warwick, Coventry, CV4 7AL, United Kingdom
\and
  Landessternwarte, Zentrum f\"ur Astronomie der Universit\"at Heidelberg, K\"onigstuhl 12, 69117 Heidelberg, Germany
\and
Institut f\"ur Astrophysik und Geophysik, Georg-August-Universit\"at G\"ottingen, Friedrich-Hund-Platz 1, 37077 G\"ottingen, Germany
\and
  Institut de Ci\`encies de l’Espai (ICE, CSIC), Campus UAB, c/ de Can Magrans s/n, 08193 Cerdanyola del Vall\`es, Barcelona, Spain
\and
  Institut d’Estudis Espacials de Catalunya (IEEC), c/ Gran Capit\`a 2--4, 08034 Barcelona, Spain
\and
Hamburger Sternwarte, Gojenbergsweg 112, 21029 Hamburg, Germany
}

 \date{Received 19 July 2024 / Accepted 21 October 2024}

 \abstract
 {}
 {
  Our goals are to characterise the kinematic properties and to identify young and old stars among the M dwarfs of the CARMENES input catalogue.}
 {We compiled the spectral types, proper motions, distances, and radial velocities for 2187 M dwarfs.
 We used the public code {\tt SteParKin} to derive their galactic space velocities and identify members in the different galactic populations. We also identified candidate members in young stellar kinematic groups, with ages ranging from 1\,Ma to 800\,Ma with {\tt SteParKin}, {\tt LACEwING}, and {\tt BANYAN}~$\Sigma$. 
 We removed known close binaries and perform an analysis of kinematic, rotation, and activity indicators (rotational periods and projected velocities, H$\alpha$, X-rays, and UV emission) for 1546 M dwarfs. 
 We defined five rotation-activity-colour relations satisfied by young ($\tau \lesssim$ 800\,Ma) stars.}
 {We identified 191 young M dwarf candidates ($\sim$12\%), 113 of which are newly recognised in this work. 
 In this young sample, there are 118 very active stars based on H$\alpha$ emission, fast rotation, and X-ray and UV emission excess. Of them, 27 have also strong magnetic fields, 9 of which are likely younger than 50\,Ma.
 Additionally, there are 87 potentially young stars and 99 stars with a dubious youth classification, which may increase the fraction of young stars to an astounding 24\%.
 Only one star out of the 2187 exhibits kinematics typical of the old Galactic halo.}
 {A combined analysis of kinematic and rotation-activity properties provides a robust method for identifying young M dwarfs from archival data.
 However, more observational efforts are needed to ascertain the true nature of numerous young star candidates in the field and, perhaps more importantly, to precisely quantify their age.}


 \maketitle
%

\section{Introduction}

\begin{table}
\centering
\caption{Ages of some representative stellar kinematic groups and associations.}
\label{table.skg}
\begin{tabular}{l c l }
 \noalign{\smallskip}
 \hline
 \hline
 \noalign{\smallskip}
SKG     &               Age             &       Reference$^a$           \\
                &               [Ma]    &                                       \\
\noalign{\smallskip}
 \hline
\noalign{\smallskip}
Taurus-Auriga           &       1--10           &       KH95, Luh04             \\ 
Upper Scorpius  & $\sim$ 10 &  PM16 \\
Argus                           &       $\sim$ 40       & Torr08        \\ 
IC 2391 Supercluster            &       $\sim 50$       &       Barr04  \\ 
$\alpha$ Persei &  $\sim 90$    &       Sta99 \\
Pleiades  &  $\sim 115$ &  Bas96  \\
Local Association       &       10--300         &                       \\ 
~~~~{TW Hydrae}                 & $\sim 10$     &       Bell15          \\ 
~~~~{$\eta$ Chamaeleontis}      & 10--15        &       Mam99, Bell15   \\ 
~~~~{$\beta$ Pictoris}  & $\sim 25$     &       Mess16          \\ 
~~~~{Columba}   &       $\sim 40$       & Bell15                \\ 
~~~~{Carina}    &       $\sim 40$       & Bell15                \\ 
~~~~{Tucana-Horologium} & $\sim 50$     &       Bell15          \\ 
~~~~{AB Doradus}        &       $\sim 150$      & Bell15                \\ 
Hercules-Lyra                                   &       200--300 &              Eis13   \\ 
Ursa Major                              &       $\gtrsim 300$   &       Gia79, SM93            \\ 
Castor                          &       $\gtrsim 300$   &       Barr98          \\ 
Hyades open cluster                             &       $600- 830$      & Dou19   \\ 
\noalign{\smallskip}
\hline
\end{tabular}
\tablefoot{
\tablefoottext{a}{Barr98: \cite{Barrado98}; Barr04: \cite{Barrado04}; Bas96: \cite{Basri96}; Bell15: \cite{Bell15}; Dou19: Table 1 in \cite{Dou19} Eis13: \cite{Eisenbeiss13}; Gia79: \cite{Giannuzzi79}; KH95: \cite{KH95}; Luh04: \cite{Luhman04}; Mam99: \cite{Mamajek99}; Mess16: \cite{Mess16}; PM16: \cite{Pecaut16}; Perr98: \cite{Perryman98}; SM93: \cite{SM93}; Sta99: \cite{sta99}; Torr08: \cite{Torr08}.}
}
\end{table}

M dwarfs make up over two-thirds of the main sequence population in the solar neighbourhood \citep{Hen06,Boch10,Reyle21}.
Because of their abundance, they are excellent targets for broad studies of the formation and evolution processes of stellar objects at the bottom of the main sequence in the Hertzsprung-Russell diagram \citep[][and references therein]{Chabrier03}.
In addition, M dwarfs are prime targets for planet surveys. 
Their low masses and small radii (between 0.6 and $\sim$0.1\,M$_\odot$, and 0.6 and $\sim$0.1\,R$_\odot$ -- \citealt{Reid95,Schw19}) represent a huge advantage with respect to solar-like stars for the detection of potentially habitable Earth-sized planets with the radial-velocity \citep[e.g.][]{Mayor09, Bonfils13, Wright16,Zech19,Dreizler19,Dreizler24} and transit \citep[e.g.][]{BT15,Gillon16,Ditt17,Luque19,Trif21} methods.
The characteristics of the host M dwarfs, in terms of luminosity or flaring activity, and of the planet parameters, such as semi-major axis, bulk density, and tidal locking strongly influence the habitability of the detected planets \citep{Scalo07, Tarter07}. 
Although M dwarfs have been the subject of numerous publications that determine their astrophysical parameters (from \citealt{AJ22} and \citealt{Luyten22} to \citealt{Reiners2022} and \citealt{Kirk24}  to offer a mix of prior and recent examples),  their precise ages have not yet been determined \citep{Upgren78,Fleming95,Gizis2002,KS07,Kiman21,Pop21,Engle23}.
Early M dwarfs have convective envelopes and radiative cores that become smaller and disappear altogether at spectral
types of $\sim$M3--4 and later. 
The generation of magnetic flux arise from dynamo actions, which are different for stars with radiative cores and for fully convective stars \citep[e.g.][]{Browning08}.
The effect of this magnetic field is observed, for example, through the emission of the H$\alpha$ line at 6562.8\,$\AA$ originated in the stellar chromosphere \citep{Hawley96,West04,Jeff18,Schf19}. The generation of magnetic flux is closely connected to the stellar corona, where the heated plasma is trapped and produces radiation in the X-ray domain \citep{Gudel04}.
There is also an intimate relation of chromospheric and coronal activity with rotation \citep{Noyes84,Pizzolato03,Wright11}, with rapid rotators showing higher levels of magnetic activity \citep[e.g.][]{Stelzer16}.
Additionally, there is a direct relation between coronal and chromospheric emission with ultraviolet (UV) emission \citep{Walkowics09, Stelzer13, Pineda20}.

Stellar activity is well-connected with age: while very young stars exhibit strong signatures of activity \citep[e.g.][]{Joy45,Bertout88,PF05}, this activity decays with increasing age \citep[e.g.][]{Silvestri05, Mamajek08, West08,Zhao11, Davenport19}. 
The tight relation between activity and age provides constraints to searches for young stars \citep[e.g.][]{Torr06,Rodriguez13,Bell15,Binks15a}, which are typically found in streams of stars travelling together, with similar galactic spatial velocities ($UVW$) and presumably sharing the same origin \citep[e.g.][]{Montes01,Malo13}.
These streams are stellar kinematic groups (SKGs) and are relatively young (under 1\,Ga).
In the solar vicinity, there are over a dozen young SKGs with ages ranging from $\sim$10\,Ma (such as TW Hydrae) to over $\sim$800\,Ma in the case of the Hyades supercluster, which are believed to share age with the open cluster (\citealt{BH15}; see Table~\ref{table.skg}). However, their ages are not always well constrained and lists of group members are not free of older stellar contaminants \citep{Tabernero12, Mitschang13}.
In particular, the presence of low-mass star members in SKGs provides crucial knowledge on their formation and evolution. These properties, in turn, have an important influence on the evolution of the Galaxy on timescales longer than the current age of the Universe \citep{Laughlin97}.

The identification and characterisation of new components of these SKGs are recurrent topics of research, from individual research groups \citep[e.g.][]{King03, Filippazzo15, Aller16} to large projects such as the Search for Associations containing Young Stars  (SACY; e.g. \citealt{Torr06,Elliott16}) or the All-sky Co-moving Recovery Of Nearby Young Members (ACRONYM; \citealt{Shk17, Schn19}). In parallel, some groups have focussed on developing algorithms and methodologies for identifying new members, such as \cite{Montes01}, \cite{Zuck04}, or \cite{Ammler16}, and for assessing membership, such as the series of papers presenting the Bayesian Analysis for Nearby Young AssociatioNs ({\tt BANYAN}, from \citealt{Malo13} to \citealt{Gagne18XIII}) and  LocAting Constituent mEmbers In Nearby Groups ({\tt LACEwING}, \citealt{Ried17Lacewing}) codes.
There are also some studies devoted to characterising the local galactic structure that result in the identification of new moving groups, such as Melange-3 \citep{Barber22}, 32~Ori \citep{Luhman22}, and Oceanus \citep{Gagne23}, among others \citep[for example,][]{Oh17,KC19}.

We note several previous papers that focus on kinematics and youth of low- and very low-mass stars in the following. 
\cite{RB09}  analysed the Li~{\sc i} 6708\,{\AA} line in high-resolution spectra, determined an average kinematic age of a sample of M7--8.5 dwarfs at 3.1\,Ga, and also found and dated some young brown dwarfs. 
\cite{Shk12}  studied a sample of K7--L0 dwarfs younger than 300\,Ma and suggested memberships in different stellar kinematic groups for each component. 
\cite{Shk17} confirmed K7--M9 members of $\beta$~Pictoris moving group and photometrically identified new candidate members. In addition, \cite{Schn19} confirmed new young K5--M5.5 moving-group members based on kinematics and age diagnostics, such as  the H$\alpha$ and Li~{\sc i} equivalent widths, X-ray and UV fluxes, and colour-magnitude diagram positions.

Using their kinematics, stars can also be assigned to the different galactic stellar populations, which are characterised by different scale heights: $\sim$100\,pc for the young disc \citep{Ng97}, $\sim$200\,pc for the thin disc \citep{CL07}, 800--1050\,pc for the thick disc \citep{Buser99,CL07}, and 2.7--3.2\,kpc for the halo \citep{Sandage87,Layden95}. Additionally, \cite{Leg92} delimited the region where young stars lie within the thin disc of the Galaxy. In a similar manner, stars belonging to each population can be identified, since their galactocentric velocities show distinguishable velocity dispersions \citep[see][]{Bensby03,Bensby05}.
In addition, the metallicity content varies from low heavy element and iron abundances in thick-disc and halo stars to near solar composition in thin-disc stars \citep{Gilmore85}. This opened the door to chemical tagging \citep{DeSilva07,Hawkins15,Marino19}.
The assignation to a population thus implies a rough age estimation, with thin-disc stars being younger than $\sim$8\,Ga \citep{Fuhrmann98} and halo stars being as old as the age of the Galaxy ($\sim$13\,Ga; \citealt{Cayrel01}).

This is the eighth paper of the series of publications on the CARMENES (Calar Alto high-Resolution search for M dwarfs with Exoearths with Near-infrared and optical Echelle Spectrographs) input catalogue of M dwarfs, aimed at identifying young and old stars in the database based on their activity, rotation, and kinematics, as well as providing a robust observational methodology for carrying out such a task in the M-dwarf regime. 
The dual-channel CARMENES spectrograph\footnote{\url{http://carmenes.caha.es}}, which started operating in early 2016, is dedicated to the detection and characterisation of Earth-like planets in the habitable zone of low-mass stars in the solar vicinity with precise radial velocities \citep{Quirr14,Quirr20}. 
So far, CARMENES has targeted slightly over 400 of the nearest, brightest, M dwarfs observable from southern Spain and discovered more than 70 exoplanets only as part of the guaranteed time observations (see \citealt{Ribas23}).

The paper is structured as follows. 
In Sect. 2 we present the sample coordinates, distances, proper motions, radial velocities, and activity-related properties (X-rays, UV, and H$\alpha$ emission, and rotational period and velocity).
In Sect.~\ref{sec.analysis} we describe the procedure followed for the kinematic analysis and define the activity-spectral type relations.
In Sect.~\ref{sec.discussion}, we present and discuss the results. In Sect.~\ref{sec.conclusions} we provide a summary of the work.


\section{Data}

\subsection{Sample, spectral types, and coordinates}\label{sec.sample}

\begin{figure*}
\centering
\includegraphics[width=0.33\textwidth]{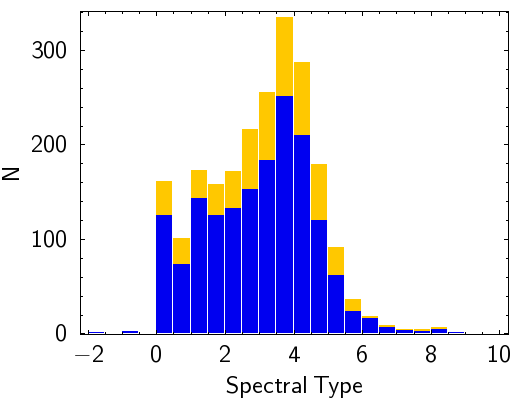}
\includegraphics[width=0.33\textwidth]{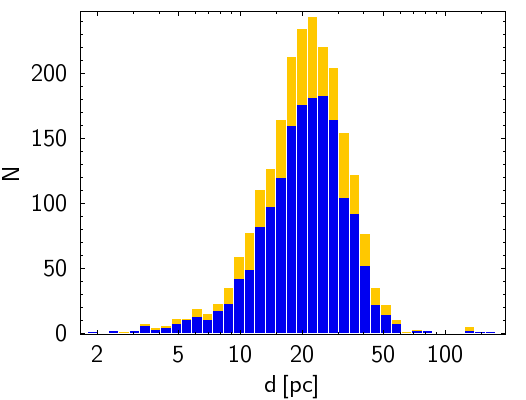}
\includegraphics[width=0.33\textwidth]{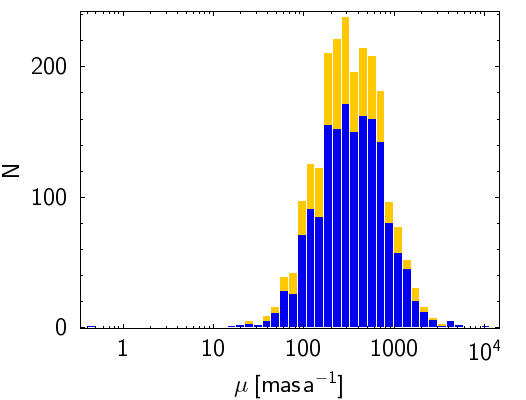}
\caption{Spectral type ({\em left}), distance ({\em middle}), and total proper motion ({\em right}) distributions of stars in the Carmencita catalogue. 
K5 and K7\,V stars are represented with $-$2 and $-$1, respectively, and M0.0 to M9.0\,V stars with numbers from 0.0 to 9.0. 
In the three panels, blue bars correspond to the stars with radial velocities for which we perform the kinematic and youth analysis, and yellow bars to the remaining Carmencita stars with known stellar companions at less than 5\,arcsec.} 
\label{fig.hist_3} 
\end{figure*}

\begin{figure}
\centering
\includegraphics[width=\hsize]{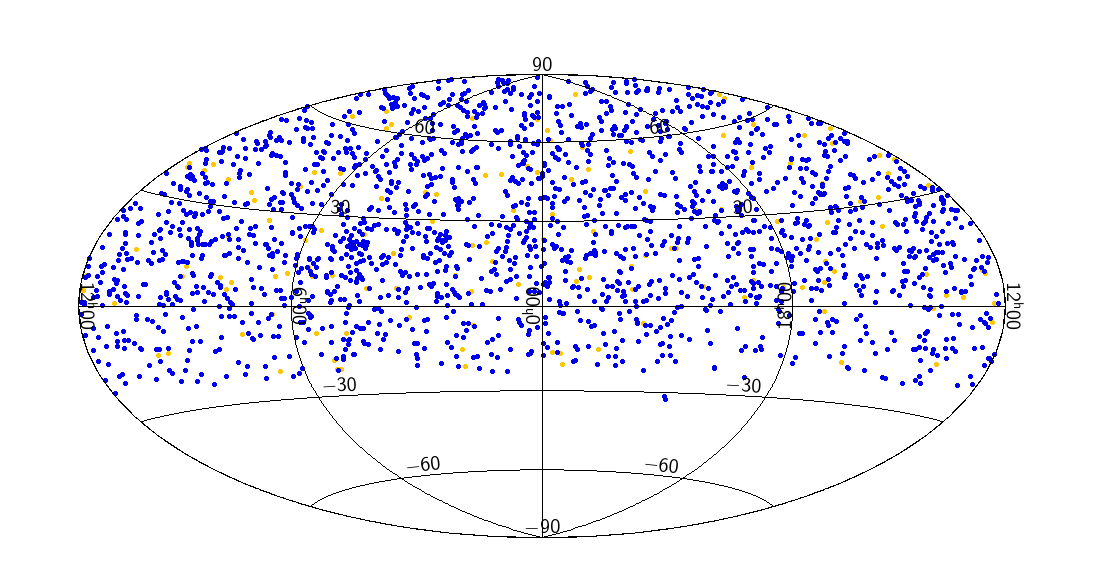}
\caption{Sky distribution in Aitoff projection of stars in the Carmencita sample.
Same colour legend as in Fig.~\ref{fig.hist_3}.}
\label{fig.skydist}
\end{figure}

Our initial sample was ``Carmencita'' \citep[CARMENes Cool dwarf Information and daTa Archive --][]{cab13,2016csss.confE.148C,Quirr15,AF15}, the input catalogue of nearby, bright M dwarfs from which the targets for radial-velocity monitoring by the CARMENES exoplanet survey were selected.

Our initial sample was taken from Carmencita, the input catalogue of nearby, bright M dwarfs from which the targets for radial-velocity monitoring by the CARMENES exoplanet survey were selected \citep{cab13,2016csss.confE.148C,Quirr15,AF15}.
CARMENES data have also been devoted to the characterisation of M dwarfs through the extensive analyses carried out to study stellar \citep{Schw19,Cif20,BelloG23} and atmospheric \citep{Pass20,Marfil21} parameter determination, abundances \citep{Abia20,Shan21,Tabernero24}, rotation \citep{DA19,Fuhr19,Shan24}, and activity \citep[e.g.]  []{Schf19,Schf22,Lafarga20,Fuhr23}. 
Further investigations on star-planet interaction, occurrence rates of exoplanets around M dwarfs, multiplicity, and magnetic fields (amongst other properties) are also underway.

During the last decade, we have carefully characterised over 2000 M dwarfs in the CARMENES input catalogue with low- and high-resolution spectroscopic \citep{AF15, Jeff18,Perd21}, imaging \citep{Cor17}, photometric (\citealt{DA19, Cif20}), and astrometric (this work) surveys.
Since the CARMENES survey is on-going, Carmencita is in constant evolution.
At the time of starting this study (March 2023), it contained dozens of stellar parameters for 2218 late-type dwarfs; all of them are in the spectral type range from M0 to M9, except for 3 K7 dwarfs and 1 K5 dwarf.
Half of these spectral types come from \citet{Hawley96}, followed by the determinations by \citet{Lep13} and \citet{AF15} with near one-fifth each; the remaining $\sim$9\% of spectral types were recovered from various sources.
The distribution of spectral types, shown in the left panel of Fig.~\ref{fig.hist_3}, ranges from K5\,V to M9.0\,V, with a peak at M3.5\,V.
However, most of the Carmencita stars have spectral types M4.5\,V or earlier because CARMENES is capable of measuring precise radial velocities only on targets brighter than $J$ = 10.5\,mag.

By construction, all Carmencita stars have an entry in the 2MASS catalogue (Two Micron All Sky Survey, \citealt{Skru06}). However, we used right ascension and declination from the third data release (DR3) of {\em Gaia} \citep{GaiaBrown,GaiaDR3}. 
Their declinations are above $-$23\,deg (observable from Calar Alto with airmasses lower than 2.0), except for AU~Mic and its companion AT~Mic~AB \citep{Cab09,Cale21}.
The sky distribution of the whole sample is shown in Fig.~\ref{fig.skydist}.
We present in Table~\ref{table.name_coords_pm_d} the 2218 stars of the Carmencita catalogue.
In the first seven columns, we provide their CARMENES identifier \citep[``Karmn'' --][]{Quirr15, AF15}, common or discovery name, {\em Gaia} DR3 source designation, spectral type with its reference, and equatorial coordinates.

Finally, we emphasise that the criteria for inclusion of a star in Carmencita are just spectral type, $J$-band magnitude, and declination.
Besides, the M dwarfs eventually monitored by CARMENES are just the brightest ones for their spectral type with just one additional selection criterion: no companions at $\rho \lesssim$ 5\,arcsec \citep{Cor17,Rein18}.
As a result, there is no bias in our sample based on rotational velocity, metallicity, or distance; although there may remain a Malmquist bias towards younger, intrinsically brighter stars.

\subsection{Distances}

Distances for 2189 K and M dwarfs in Carmencita, representing 98.7\% of the sample, come from trigonometric parallax determinations, mainly from {\em Gaia} DR3. If not available in {\em Gaia} DR3, we took parallaxes from {\em Gaia} DR2 \citep{GaiaDR2} or other sources instead.
There are nine close binaries in triple systems for which {\em Gaia} DR3 does not provide astrometric solutions; we used the solutions from the primary component instead.
Another two stars belong to the Taurus-Auriga association, for which we adopted the distance to the dark cloud \object{Lynds~1495} from \cite{Galli18}. 
The remaining 27 stars (1.3\%) have spectophometric distances derived as by \cite{Cor17}, eleven of which should be considered as a lower limit due to the presence of a companion at less than 5\,arcsec that could contaminate the photometry used for the calculations. 
Close binaries in Carmencita were either identified by us \citep{Cor17,Jeff18,Baro18,Baro21} or by other authors \citep[e.g.][]{Jan12, Jan14, Jodar13, Bowler15palmsV}. 
An extensive analysis of multiplicity in Carmencita will be published in \citet{Cifuentes24}.
The full list of distance references and the corresponding number of stars are listed in the top part of Table~\ref{table.3_ref}.

We provide compiled and estimated distances, with errors and references, for all the 2218 Carmencita stars in  columns 11 and 12 of Table~\ref{table.name_coords_pm_d}.
The middle panel in Fig.~\ref{fig.hist_3} shows the distribution of distances in the sample. 
They range from 1.8\,pc (\object{Barnard's Star} / J17578+046) to 166.1\,pc (\object{Haro 6--36} / J04433+296).
The median distance is 21.3\,pc and almost the entire catalogue (99\%) lies within 55\,pc from the Sun. 
There are seven M dwarfs farther than 100\,pc, all of which are young and overluminous \citep{Cif20}: five in Taurus-Auriga (XEST 16--045 / J04206+272, FW Tau / J04294+262, V927 Tau / J04313+241, Haro 6--36 / J04433+296, and PM J04393+3331 / J04393+335), one in Upper Scorpius (K2--33 / J16102$-$193) and the very young star StKM 1--1155 / J14259+142, which was classified as a rejected $\beta$~Pictoris candidate member by \citet{Lee24}.

\subsection{Proper motions}

{\em Gaia} DR3 \citep{GaiaDR3} -- and when data were lacking, then DR2 -- supplied 95.5\% of the proper motions gathered in Carmencita. 
For the remaining stars, we compiled proper motions mainly from the PPMXL Catalog \citep{Roeser10}, the new reduction of {\em Hipparcos} \citep{HIP2}, or our own measurements.
In particular, we improved the proper motion determination of 22 stars, some of them with PPMXL uncertainties greater than 10\,mas\,a$^{-1}$, with the methodology of \cite{Cab10}.
We used the Aladin sky atlas \citep{Bonn00, BF14} and the astrophotometric catalogues {\em Gaia},  US Naval Observatory catalogue USNO-A2.0 \citep{Monet98},  Guide Star Catalog GSC2.3 \citep{Morrison01}, 2MASS,  Carlsberg Meridian Catalog CMC14/15 \citep{Evans02},  Sloan Digital Sky Survey SDSS DR9 \citep{SDSS9},  All-sky Wide-field Infrared Survey Explorer ALLWISE catalogue \citep{Cutri14}), and (when available) Astrographic Catalogue AC2000.0 \citep{Urban98}. 
For the 22 stars, the typical time baseline between the first and last epoch was of about 60\,a and the resulting proper motion errors we computed  never exceeded 1.1\,mas\,a$^{-1}$ in each component.

The references for the adopted proper motions in Carmencita and corresponding number of stars are listed in the middle part of Table~\ref{table.3_ref}. 
The right panel in Fig.~\ref{fig.hist_3} shows the total proper motion distribution of the catalogue. 
The two stars at the edges of the distribution are Barnard's Star / J17578+046, with a total proper motion of 10\,400\,mas\,a$^{-1}$, and \object{HD~168442} / J18198$-$019, which is moving towards the Sun with a proper motion of 0.43\,mas\,a$^{-1}$ \citep{BJ18_stellarencounters}. 
The median motion of all stars is 326\,mas\,a$^{-1}$.
Columns 8-10 in Table~B.1 display the proper motions of the Carmencita stars, with errors and references.

\subsection{Radial velocities}
\label{sec.rv}

In Carmencita, there are 2187 (98.6\%) stars with radial velocities measured in the literature. They were compiled using the CDS {\tt X-Match} service\footnote{\tt http://cdsxmatch.u-strasbg.fr/} \citep{pineau11} and the Tool for OPerations on Catalogues And Tables \citep[{\tt Topcat};][]{Taylor05} or that we computed  on CARMENES, FEROS (Fiber-fed Extended Range Optical Spectrograph), CAFE (Calar Alto Fiber-fed Echelle spectrograph), or HRS (High Resolution Spectrograph) high resolution spectra \citep[e.g.][]{Jeff18,Lafarga20}. 
In this work, we provide a new radial velocity measurement for J19255+096 from CARMENES spectra \citep{Ribas23} computed via a cross-correlation function, as done by \cite{Lafarga20}.
The references and number of stars of each of them are listed in the bottom part of Table~\ref{table.3_ref}.

For consistency, we took the mean absolute radial velocities from \cite{Lafarga20}, which were not corrected for  gravitational redshift or convective blueshift.
\cite{Reid95} and \cite{Gizis2002} did not provide radial velocity uncertainties in some cases. We assigned them the typical accuracy of their catalogues: 15\,km\,s$^{-1}$ and 1\,km\,s$^{-1}$, respectively.
For radial velocities of \cite{Ding22}, we included the 5.0\,km\,s$^{-1}$ uncertainty of the pipeline as described by \cite{Xiang15}.
Errors in the radial velocities of \cite{Sou18} were computed from the quadratic sum of the radial velocity uncertainty and the standard deviation.
Adopted radial velocities, together with their references, are shown in  columns 2-4 in Table~\ref{table.name_vr_uvw}.

Of the 2187 stars with radial velocities, there are 606 with a known close physically related companion at less than 5\,arcsec that could compromise their activity, astrometric, and photometric parameters (e.g. X-rays, proper motions, and GALEX and WISE photometry; see more details below),
in addition to their kinematics.
The identification of these 606 close companions is the result of a detailed analysis of multiplicity in the Carmencita sample described by \cite{Cifuentes24}. 
They identified the companions in different manners:
($i$) a search for common proper motion and parallax pairs in \textit{Gaia} DR3;
($ii$) a selection of \textit{Gaia} DR3 multiplicity indicators (e.g. {\tt RUWE}, {\tt e\_RV}, {\tt duplicated\_source}, {\tt non\_single\_star}, {\tt phot\_variable\_flag};
($iii$) a star-by-star analysis of stars in the Washington Double Star catalogue \citep{WDS};
and ($iv$) a complementary literature search for spectroscopic binaries and very close systems not resolved by \textit{Gaia} (but with adaptive optics or speckle imaging).

As for the radial velocities, there are only 17 spectroscopic multiple systems with determinations of the barycentric radial velocity of the system ($\gamma$) by \cite{Baro18, Baro21}.
Except for one (LP~427--16 / J09140+196), they all have well determined orbital periods and, thus, their radial-velocity measurements take into account the presence of the close companion.
Another 429 close binary stars have radial velocities provided by {\em Gaia} and \cite{Hal18} that correspond to the mean values of the set of observations, and their errors correspond to the standard deviations. These radial velocities would equal the systemic velocity only in the case of systems with orbital periods shorter than the {\em Gaia} DR3 time coverage or with orbital periods that are long enough not to show radial velocity variations during the period of observations. Therefore, we cannot assume that the given radial velocities are in all 429 cases the systemic velocities.
The remaining 160 stars with close companions have radial velocities provided by other articles and correspond in most cases to single-epoch values.
We calculated galactocentric velocities and their membership in the different galactic populations for the whole sample, but this attribution should be taken with caution in the case of the 606 stars with companions at $\rho <$ 5\,arcsec. For the kinematics-activity-rotation analysis, we discarded them and kept the remaining 1581 (2187$-$606) stars.

\subsection{Rotation and activity indicators}

\begin{figure}
\centering
\includegraphics[width=1.1\hsize]{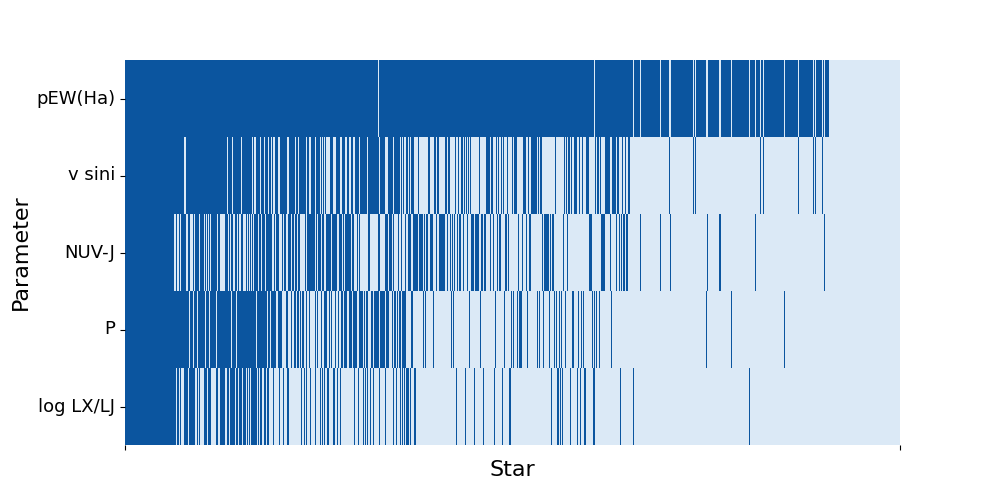}
\caption{Grid heat map of stars with (dark blue) or without (light blue) activity indicators in Table~\ref{tab.activity_num_params}.
They are sorted by the number of compiled parameters per star.
}
\label{fig.heatmap}
\end{figure}

To account for the rotation and activity level of the 1581 input stars under study, we gathered the following measurements: 
rotational periods ($P$), 
projected rotational velocities ($v\sin{i}$), 
X-ray emission (quantified by the $L_X/L_J$ quotient),
and near-ultraviolet (NUV) and near-infrared (NIR) colours from Galaxy Evolution Explorer ({\em GALEX}, \citealt{GALEX}) and 2MASS ($NUV-J$), 
as well as H$\alpha$ pseudo-equivalent widths (pEW(H$\alpha$)). 
In most cases, we compiled these parameters from the literature with the Tool for OPerations on Catalogues And Tables \citep[{\tt Topcat};][]{Taylor05}, the SIMBAD astronomical database \citep{Wenger00}, and the VizieR catalogue access tool \citep{Och00}. 
We also included our own measurements obtained from internal data and published in a number of articles \citep[e.g.][]{AF15, Jeff18,Shan24}.
The adopted values and their references are displayed in Table~\ref{table.name_activity_params}. 
In Table~\ref{tab.activity_num_params}, we summarise the number of stars among the 1581 for which we compiled kinematics-rotation-activity measurements. 
To better illustrate these values, we represent in Fig.~\ref{fig.heatmap}
a grid heat map of the parameter availability per star.
We provide more details on the compilation or measurement of each parameter below.
In Table~\ref{tab.indicatorsrefs}, we list the references used for the data compilation.

\begin{table}[]
\centering
\caption{Number of input stars with available measurements of rotation and activity indicators and no known companions at $\rho <$ 5\,arcsec.}
\label{tab.activity_num_params}
\small
\begin{tabular}{l lc }
 \noalign{\smallskip}
 \hline
 \hline
 \noalign{\smallskip}
 Parameter & Units  & $N$ \\
 \noalign{\smallskip}
 \hline
  \noalign{\smallskip}
$v_{\rm r}$ & km\,s$^{-1}$  &  1581  \\
$P$ & d  &  511 \\
$v \sin{i}$ & km\,s$^{-1}$  &  626  \\
$\log{L_X/L_J}$ &  dex &  361 \\
$NUV-J$  & mag & 566 \\
pEW(H$\alpha$) &  \AA &  1389  \\
\noalign{\smallskip}
\hline
\end{tabular}
\end{table}

\subsubsection{Rotational periods} 

We compiled $P$ from a number of references in the literature and included also some values measured by the CARMENES Consortium \citep{DA19,Shan24}, mostly with small ground telescopes and the {Transiting Exoplanet Survey Satellite} (TESS; \citealt{Ricker15_TESS}).
Additionally, we used the {\em Kepler} K2 mission \citep{Howell14_K2} and, again, TESS to measure rotational periods and complement the list of those gathered from the literature.
We took the photometry collected over TESS cycles 1 and 2, spanning from July 2018 to July 2020, along a total of 26 sectors,  extending over $\sim$27.4\,d each.
We studied a total of 316 stars with assigned TESS input catalogue (TIC) identifier and with 2\,min-cadence photometry available that corresponded to one or more sectors. 
As for K2, a total of 28 stars were observed, 25 of them in one single campaign and three in two different ones, each spanning a total of 80\,d.
Both the TESS 2\,min-cadence and K2 light curves were downloaded from the Mikulski Archive for Space Telescopes portal\footnote{\url{https://mast.stsci.edu/portal/Mashup/Clients/Mast/Portal.html}}.

\begin{figure}
\centering
\includegraphics[width=\hsize]{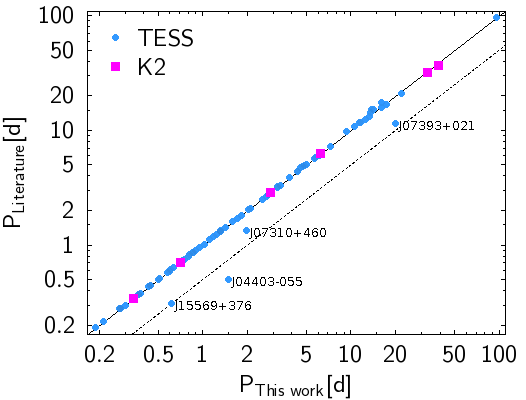}
\caption{Comparison of 84 rotational periods from the literature and measured in this work from TESS (blue circles) and K2 (magenta squares) light curves. 
Solid and dashed black lines mark the 1:1 and 1:2 relations, respectively. Four outliers are labelled.
}
\label{fig.prot}
\end{figure}

For all stars with TESS photometry, we first used the pre-search data conditioning single aperture photometry (PDCSAP) flux time series, which have had their  instrumental effects corrected in the single-aperture photometry (SAP)-only light curve.
For the stars with light curves in several TESS sectors, we stitched them without any prior normalisation. 
However, for both TESS and K2 stars, we separated the light curves into chunks; namely, a series of continuous observations with a separation of less than one day between any two adjacent points as \citet{Shan24}. We then computed the generalised Lomb-Scargle (GLS) periodograms \citep{ZK09}.
Afterwards, we folded the light curves with the periods corresponding to the most significant periodogram peaks and adopted them as the actual rotation periods of the stars if they passed a positive visual inspection. 
For stars for which we could not identify a significant rotational period using the GLS periodogram alone, we used the SAP flux time series instead. 
For stars with several chunks, we calculated and corrected the offset between the different chunks, corrected them for each phase, unfolded the light curve, and calculated a new period value from the periodogram. 
We iterated this process until the value of the period converged.
Finally, we got 137 rotational periods from TESS (116) and {K2} (21) light curves.
We estimated associated errors of 10\% of the rotational periods, in line with the associated errors discussed in detail by \cite{Shan24}.
For a few stars in only one TESS sector, we were able to determine with a high level of confidence rotational periods as long as 14.7\,d, which is about half the time span.
For stars in more than one TESS sector, we determined periods of up to 26.8\,d and of up to 39.2\,d from the K2 data. 

Of the 137 stars with period determined by us, 53 are new and 84 had previous measurements in the literature.
After an investigation of the separation and magnitude difference in the $G$-band of all \textit{Gaia} DR3 sources within 20\,arcsec to our targets, we discarded one of the new periods.
Given the large TESS pixel size, the period of 3.15\,d initially assigned to BD+45~784 / J03332+462 (M0.0\,V) actually corresponds to the K2\,V primary star HD~21845, which is located at 9.6\,arcsec and is 2.5\,mag brighter to our target.

The comparison of previous and measured periods is shown in Fig.~\ref{fig.prot}. 
There are four stars with periods that significantly differ from the literature value.
For 1RXS J073101.9+460030 / J07310+460 and RX J1556.9+3738 / J15569+376, we report periods that are nearly twice those published by \cite{Har11}, and nearly three times the one published by \citet{Skr21} for LP~655--48 / J04403$-$055.
The fourth period that does not match ours is for BD+02~1729 / J07393+021, which was extensively studied by \citet{Shan24} with SuperWASP data.
Of the 84 stars with previous measurements in the literature, we kept the values of \citet{DA19} or \citet{Shan24} in 44 cases, and replaced them by ours in the remaining 40 cases.
The resulting 92 (52+40) periods provided by us are marked as `this work' in Table~\ref{table.name_activity_params}.

\subsubsection{Projected rotational velocities}

We drew the $v\sin{i}$ values from the literature and included measurements by the CARMENES Consortium \citep[e.g.][]{Jeff18,Rein18,Reiners2022}. 
In Table~\ref{table.name_activity_params}, we indicate whether $v\sin{i}$ is an upper limit (defined by the spectral resolution of the spectrograph) or a proper value.
Uncertainties are not always available.
The 10\% of the stars with the largest velocities have $v\sin{i} >$ 20\,km\,s$^{-1}$, while more than half are very slow rotators ($v\sin{i} \leq$ 3.0\,km\,s$^{-1}$). We kept published values smaller than instrumental broadening at 1--2\,km\,s$^{-1}$ \citep{Hou10, Reiners2022}; however, in practice, they are simply slow rotators.

\begin{figure*}[]
\centering
\includegraphics[width=0.30\hsize]{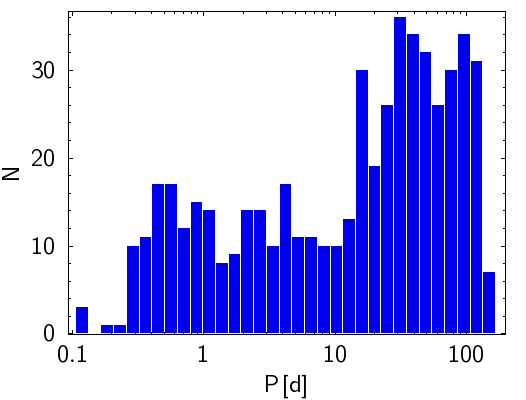}
\includegraphics[width=0.30\hsize]{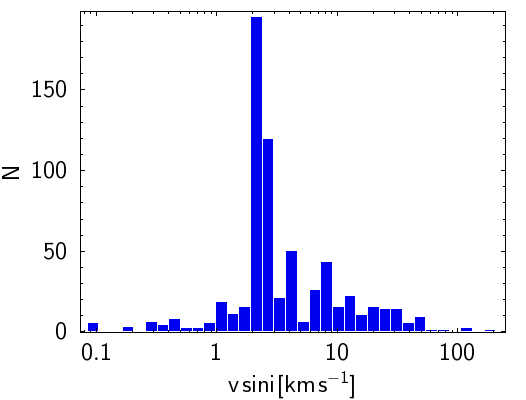}
\includegraphics[width=0.30\hsize]{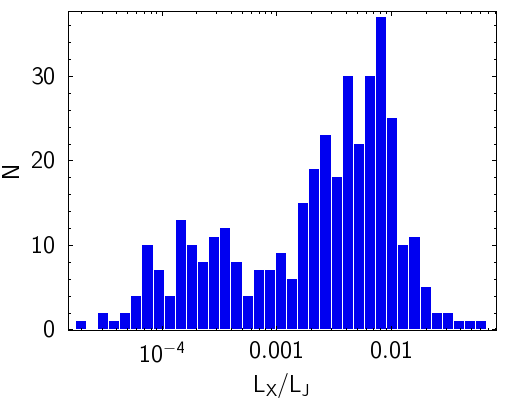}
\includegraphics[width=0.30\hsize]{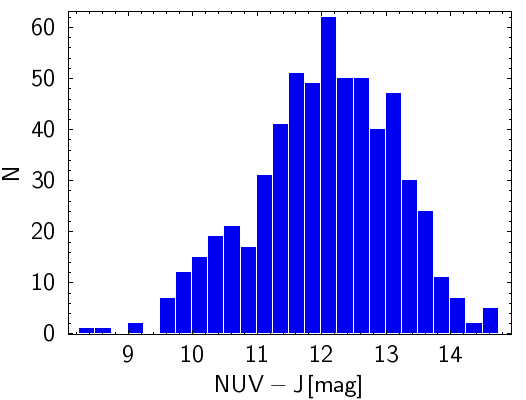}
\includegraphics[width=0.30\hsize]{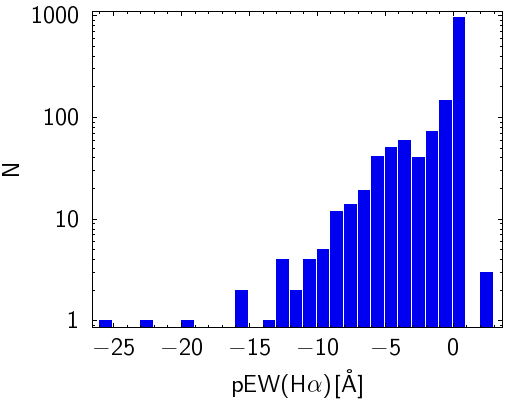}
\includegraphics[width=0.30\hsize]{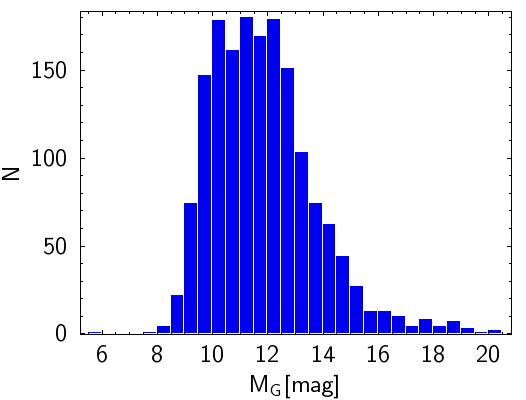}
\caption{Distribution of the rotational periods and velocities, X-rays, H$\alpha$ and UV emission, and absolute $G$ magnitude gathered for the 1581 investigated stars.}
\label{fig:hist_params}
\end{figure*}

\subsubsection{X-ray emission} 

Soft X-ray data (0.1--2.0\,keV) were taken from the {\em ROSAT} All-Sky Bright Source Catalogue \citep[1RXS,][]{Rosat}. 
The satellite was launched in June 1990 and was in operation until February 1999. 
For about two-fifths of the stars with {\em ROSAT} counterparts in this work, there was no previous association in the SIMBAD database or there was an incorrect association given with an extragalactic background source. 
It is thus the first time that these stars are linked to a {\em ROSAT} measurement.
We computed the X-ray fluxes from the source count rate and hardness ratios as \citet{Schmitt95} and, from them and the flux in the 2MASS $J$ band, the $F_{\rm X}/F_J = L_{\rm X}/L_J$ ratio.
This ratio is related to the standard $L_{\rm X}/L_{\rm bol}$ ratio through:

\begin{equation}
\frac{L_{\rm X}}{L_{\rm bol}} = \frac{L_{\rm X}}{L_J} \cdot \frac{L_J}{L_{\rm bol}} = \frac{L_{\rm X}}{L_J} 10^\frac{BC_J}{2.5},
\label{eq.lxljlbol}
\end{equation}

\noindent where $BC_J = M_{\rm bol} - M_J$ is the bolometric correction in the $J$ band. In its turn, $M_{\rm bol}$ is defined as:

\begin{equation}
M_{\rm bol} = -2.5\log{\frac{L_{\rm bol}}{L_0}} = -2.5 \log{L_{\rm bol}} + 71.197425\,{\rm mag},
\label{eq.mbol}
\end{equation}

\noindent where the $L_0$ zero-point value was described by \citet{Mamajek15_zeropoint}.
We used $L_{\rm X}/L_J$ in our analysis because $L_J$ can be be more easily computed than $L_{\rm bol}$ or $BC_J$, and is free of systematics \citep{Cif20}.

\subsubsection{$NUV-J$ colour} 

We used NUV photometry ($\Delta \lambda$ = 1771--2831\,\AA) from the {\em GALEX} space mission.
Although the International Ultraviolet Explorer \citep{IUE78} and the \textit{Hubble} Space Telescope with the Cosmic Origins Spectrograph \citep{cos_hst} operate in the same regime, {\em GALEX} covers  a much larger sky area by far.
We took the near-infrared photometry from 2MASS as the photospheric reference.
As mentioned before (and discussed extensively by \citealt{Cif20}), all Carmencita stars have a 2MASS $J$ magnitude, but not all of them have {\em GALEX} $NUV$ photometry.
As for the $NUV-G_{RP}$ colour, $NUV-J$ is also sensitive to both spectral types and especially the UV flux excess, which may be caused by chromospheric activity or interaction between close binaries \citep[cf.][and references therein]{Cif20}.

\subsubsection{H$\alpha$ pseudo-equivalent widths} 

The three major contributors to the compilation of pEW(H$\alpha$) were \cite{Gai14}, \cite{Fuhr22}, and \cite{AF15}, which supply 78\% of the data. Whenever these authors did not provide pEW(H$\alpha$), we took them from other available references by prioritising the most recent measurement and with the highest spectral resolution \citep[e.g.][]{Schf15,New17,Lu19,Schf19}.
In some cases, the selected pEW(H$\alpha$) is an average of a number of measurements over several years (e.g. from CARMENES data -- \citealt{Fuhr22}).
As is customary, a negative value indicates emission and a positive one does absorption.
The use of the pseudo-equivalent width (pEW) instead of the equivalent width (EW) in M dwarfs is due to the difficulty in (or, more accurately, impossibility of) measuring the continuum of the spectra, which leads to the measure of the pseudo-continuum and, hence, the measure of the pseudo-equivalent width. 
The emission of the H$\alpha$ line is a measurable feature of stellar magnetic activity in solar-like stars and binaries \citep[e.g.][]{Herbig85,CR89,Strassmeier90,Montes95,Lyra05,Gomes14}, but it is truly in the low-mass regime where it becomes a strong and reliable activity indicator \citep{GL86,Young89,West04, New17, Fuhr18,Schf19, Lafa21}.

\subsubsection{Distributions}

Figure~\ref{fig:hist_params} shows the distributions of each of the five indicators described above, together with the distribution of the $G$-band absolute magnitude, $M_G$
The median rotation period is 20.7\,d, and the $v\sin{i}$ distribution peak observed at 2--3\,km\,s$^{-1}$ is related to the upper limits of \citet{Jeff18} and \citet{Rein18}. 
The $\log{L_{\rm X}/L_J}$ ratio distribution shows two peaks near $-$2.2 and $-$3.6.
Regarding H$\alpha$ emission, only between 21 and 25\% of the sample have values under $-$0.75\,{\AA} and $-$0.30\,{\AA}, respectively, which are thresholds used in the literature for identifying H$\alpha$ active M dwarfs \citep{West15,Schf19}.

\section{Analysis}\label{sec.analysis}

\subsection{Galactocentric space velocities}\label{sec.galvel}

For the 2187 stars with radial velocities (including the 606 known close binaries), we computed the $UVW$ galactic space velocities. We used the {\em Gaia} equatorial coordinates ($\alpha$, $\delta$), distances ($d$), radial velocities ($\gamma \equiv v_{\rm r}$), and proper motion components ($\mu_\alpha\cos{\delta}, \mu_\delta$) as an input for the {\tt SteParKin}\footnote{\url{https://github.com/dmontesg/SteParKin}} code, which is an updated version of the {Fortran} code presented by \citet{Montes01} adapted to {Python3}.
{\tt SteParKin} requires J2000 coordinates in the International Celestial Reference System (ICRS) and does not correct for the Sun's motion.
The {\tt SteParKin} computation of $UVW$ follows

\begin{equation}
\left[ \begin{array}{c} U \\ V \\ W \end{array} \right] = \mathcal{B} \cdot \left[ \begin{array}{c} v_{\rm r} \\ { {k \cdot d \cdot \mu_\alpha\cos{\delta}}} \\ {k \cdot d \cdot \mu_\delta} \end{array} \right],
\label{eq.uvw}
\end{equation}

\noindent where $k=4.74057$\,km\,s$^{-1}$ and $\bf \mathcal{B}$ is a $3 \times 3$ matrix that is the result of combining the coordinate matrix of the star given in equatorial coordinates ($\bf \mathcal{A}$) with various rotation matrices that convert the former to the galactic coordinate system that can be described in a single matrix $\bf \mathcal{T}$ (see \citealt{JS87}):

\begin{equation}
    \mathcal{B}=\mathcal{T} \cdot \mathcal{A},
\end{equation}

\noindent where $\bf \mathcal{A}$ and $\bf \mathcal{T}$ are the following matrices and $\alpha$ and $\delta$ are the equatorial coordinates:

\begin{equation}
\mathcal{A} =  \left[ \begin{array}{ccc} +\cos{\alpha} \cos{\delta}     &       -\sin{\alpha}   &       -\cos{\alpha} \sin{\delta}    \\      +\sin{\alpha} \cos{\delta}      &       +\cos{\alpha}   &       -\sin{\alpha} \sin{\delta}    \\      +\sin{\delta}   &       0       &       +\cos{\delta} \end{array} \right]
\end{equation}

\noindent and

\begin{equation}
\mathcal{T} = \left[ \begin{array}{ccc} -0.06699 &      -0.87276        &       -0.48354        \\      +0.49273        &       -0.45035        &       +0.74458         \\      -0.86760        &       -0.18837        &       +0.46020 \end{array} \right].
\end{equation}

The errors in $UVW$ were calculated using Eq.~2 of \cite{JS87}.  In this approach, it is assumed that the errors in kinematic parameters are not correlated.
The computed galactic space velocities of our sample and their errors are provided in columns 5-10 in Table~\ref{table.name_vr_uvw}.

\subsection{Membership in stellar populations and SKGs}{\label{sec.pop}}

With the computed galactocentric velocities, {\tt SteParKin} assigns the stars to the different galactic populations, namely, the thin disc, thick disc, transition between thin and thick discs, and halo following the probabilistic approach described by \citet{Bensby03,Bensby05}.
{\tt SteParKin} also includes the definition by \cite{Leg92}, based on the results of \cite{Eggen89}, of the young disc population that lies within the thin disc, which is $-50 < U < +20$, $-30 < V < 0$, and $-25 < W < +10$ in km\,s$^{-1}$. 

\begin{figure}
\centering
\includegraphics[width=\hsize]{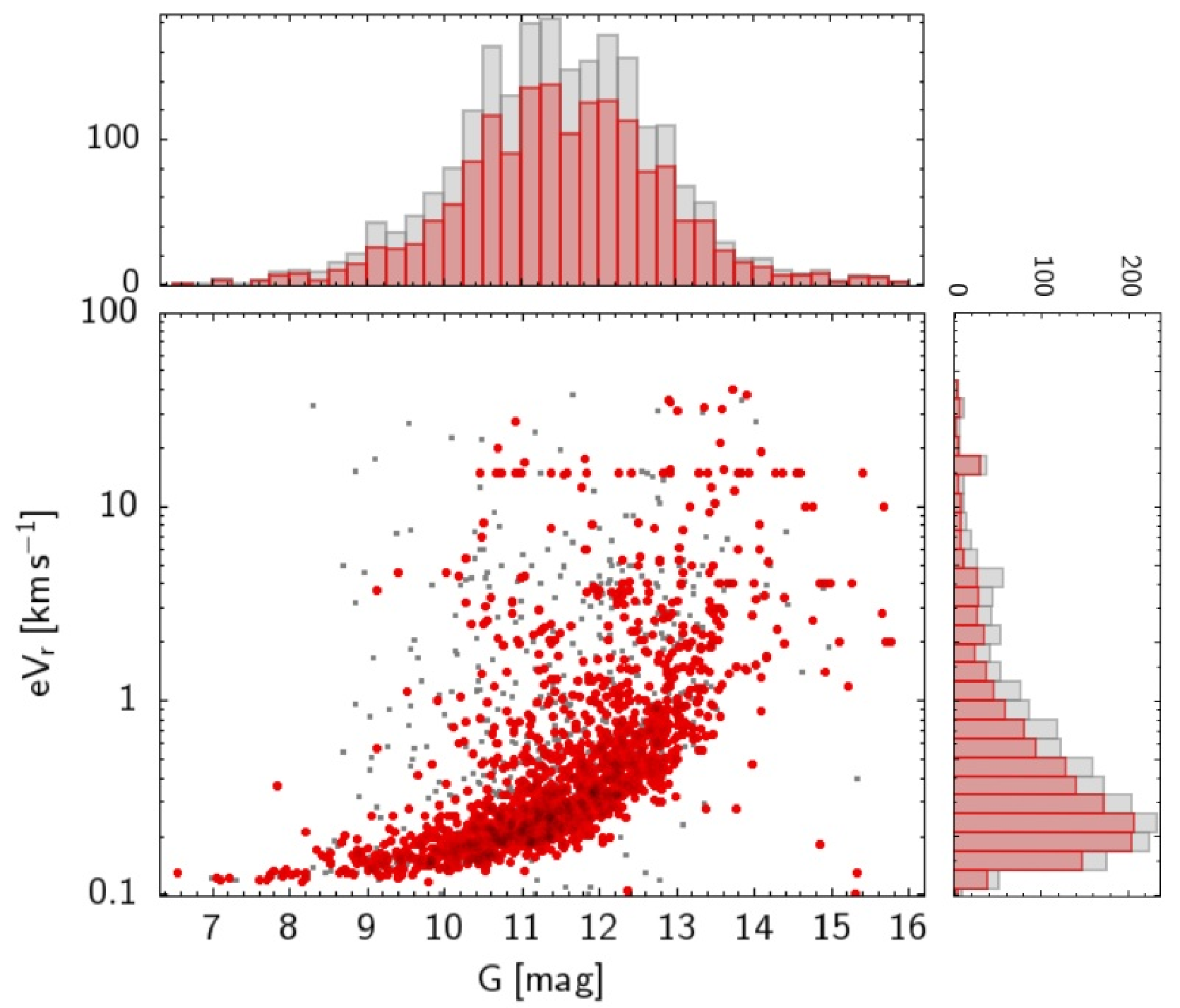}
\caption{Radial velocity errors as a function of the $G$ band magnitude for all the 2187 Carmencita stars (gray) and the 1581 single stars subject of the activity analysis (red).}
\label{fig.evr_g}
\end{figure}

The total number of young disc stars inferred from these criteria should be considered as an approximate value.
On one hand, some young stars lie slightly outside these boundaries, such as some members of the Ursa Major moving group \citep[e.g.][]{Montes01, Tabernero17}.
On the other hand, there are stellar interlopers of the thick and transition discs that have kinematics of thin and young disc stars \citep{LS06,Malo13,Antoja18}.
The errors in the radial velocities also contribute to the uncertainties in the computed $UVW$ velocities. 
In Fig.~\ref{fig.evr_g}, we represent radial velocity errors, $eV_r$, as a function of $G$-band magnitude to evaluate their significance in our results. 
Despite the large spread in the radial velocity errors, the distribution peaks at around 0.2\,km\,s$^{-1}$ and introduces an average error of 1.5\,km\,s$^{-1}$ in the Carmencita sample (and of 1.2\,km\,s$^{-1}$ in the sample of single M dwarfs). 
Therefore, we assessed the calculated galactocentric velocities used for the kinematic analysis and determined them to be overall certain. We attributed any potential misclassifications in the different age groups to other sources of error, such as the constraints applied in defining the galactic populations or SKGs, inaccuracies in the adopted measurements or the possible presence of undetected binary companions. Misclassifications due to large uncertainties in the radial velocities are considered negligible.
For a joint evaluation with these assignments, we computed the tangential velocities $v_{\rm tan} [{\rm km}\,{\rm s}^{-1}] = 4.74 \cdot d\,[{\rm pc}] \cdot \mu\,[{\rm arcsec\,a^{-1}}]$ of the stars.

We also identified stars as candidate members of different SKGs combining the outputs of three different codes: {\tt SteParKin}, {\tt LACEwING} \citep{Ried17Lacewing}, and {\tt BANYAN}~$\Sigma$ \citep{Gagne18Banyan}. 
{\tt SteParKin} assesses the membership in 15 young ($\tau <$ 1\,Ga) kinematic moving groups and associations \citep{Montes01}. 
It evaluates the galactocentric velocities lying within the triaxial ellipsoid that represents each group in the $UVW$ space. 
{\tt LACEwING} is a convergence-style algorithm that transforms goodness-of-fit metrics into membership probabilities. This code considers 13 moving groups and 3 open clusters within 100\,pc from the Sun. 
Finally, {\tt BANYAN}~$\Sigma$ is a {Python}3-based Bayesian classification algorithm that is the latest step in the refinement of the {\tt BANYAN} codes \citep[{\tt BANYAN I} and {\tt II} --][]{Malo13,Gagne14BanyanII}. 
It computes the probability of a star belonging to any of the {32} stellar associations considered within 150\,pc from the Sun or to the galactic field neighbourhood within 300\,pc. 
Both {\tt LACEwING} and {\tt BANYAN}~$\Sigma$ use the galactic position along with the galactocentric velocities of the star ($XYZUVW$) to evaluate the relevant memberships, while {\tt SteParKin} only uses the kinematic information ($UVW$).
Spatial positions not being incorporated in {\tt SteParKin} has an impact when assessing membership of SKGs that have spread over hundreds of parsecs but are still kinematically coherent, as valid members would not be left out in the case of SKGs with very large interstellar separations ($s \sim$ 100\,pc).

Futhermore, {\tt SteParKin} provides membership to the Hyades supercluster \citep{Eggen60,deSilva11,Tabernero12},
while {\tt LACEwING} and {\tt BANYAN} $\Sigma$ provide membership to the Hyades cluster \citep{vanBueren52,Hanson75,Stern95,Perryman98,Reino18}.
Since each code has its own definitions of SKGs and associations, we refer to the corresponding references for more details.

The kinematic compatibility with a SKG or association is one condition to claim probable membership, but it must be accompanied by additional photometric and spectroscopic criteria for a solid membership assignment. 
We include a complementary analysis of stellar rotation and activity signatures to assess the youth of the stars, as explained in the following.

\subsection{Kinematics and activity}
\label{sec.results.activity}

Given that stellar rotation and activity vary with the spectral type and that they decay with age, we would expect to differentiate young sources (as defined below) along the M sequence in the parameter-spectral type diagrams.
To take into account the evolution of rotation-activity with age and spectral type within the M-dwarf regime, we defined upper or lower envelopes of our parameters as a function of $G-J$ colour from a set of nearby comparison stars with well determined ages.
First, we compiled a comprehensive list of FGKM stars in the roughly coeval Hyades and Praesepe clusters (600$-$800\,Ma -- \citealt{Hernandez98,Perryman98,Delo11,Brandt15,Rebull17Praesepe,Gossage18}).
We collected 941 and 567 stars in the Praesepe and Hyades clusters from \citet{Rebull17Praesepe} and \citet{Dou14}, respectively.
For these stars, there are homogeneous pEW(H$\alpha$) and rotational period determinations by \citet{Dou14} and \citet{Dou19}. X-ray data for both clusters by \citet{Nunez22} were used to convert the X-ray count rates and hardness ratios to $L_{\rm X}/L_J$. We computed $NUV-J$ after cross-matching with 2MASS and GALEX as for our CARMENES M dwarfs.
For retrieving the $v\sin{i}$ of cluster members, we made use of the Spanish Virtual Observatory (SVO) Discovery Tool\footnote{\url{http://sdc.cab.inta-csic.es/SVODiscoveryTool/jsp/searchform.jsp}}, which is a simple user interface that allows for rotational velocities and many other parameters to be drawn from VizieR catalogues, using target names or equatorial coordinates as input. 
They mainly come from \cite{Abdurro_apogee} with APOGEE-2 data, but also from other publications such as \cite{Mermilliod09}, \cite{Kovaks18}, and \cite{Luo_vsini}.

\begin{table}[]
\centering
\caption{Bona fide members of young SKGs.}
\label{tab.55feten}
\small
\begin{tabular}{lll}
\hline
\hline
\noalign{\smallskip}
Karmn$^a$       &       Name    &       SKG     \\

\noalign{\smallskip}
\hline
\noalign{\smallskip}
\multicolumn{3}{c}{\includegraphics[width=2mm]{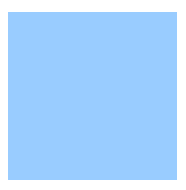} Age $<$ 50\,Ma} \\
\noalign{\smallskip}
J02519+224*  &  RBS 365  &  $\beta$ Pic \\
J03242+237  &  LP 356-15  &  IC 2391/Argus  \\
J03275+222*  &  J0327+2212  &  $\beta$ Pic  \\
J04206+272*  &  XEST 16-045  &  Tau   \\
J04472+206  &  RX J0447.2+2038  &  IC 2391/Argus  \\
J04595+017*  &  V1005 Ori  &  $\beta$ Pic  \\
J05019+011*  &  1RXS J050156.7+010845  &  $\beta$ Pic  \\
J05062+046*  &  RX J0506.2+0439  &  $\beta$ Pic  \\
J05339$-$023*  &  RX J0534.0-0221  &  $\beta$ Pic   \\
J06075+472*  &  LSPM J0607+4712  &  Columba \\
J07446+035*  &  YZ CMi  &  IC 2391/Argus  \\
J07467+574  &  G 193-065  &  Columba  \\
J09449$-$123* &  G 161-071  &  Argus  \\
J16102$-$193*  &  K2-33  &  USco \\
J19511+464  &  G 208-042  &  IC 2391/Argus \\
J20435+240*  &  Wolf 1360  &  $\beta$ Pic \\ 
J20451$-$313*  &  AU Mic  &  $\beta$ Pic  \\ 
J21100$-$193*  &  BPS CS 22898-0065  &  $\beta$ Pic \\ 
J22088+117*  &  2MASS J22085034+1144131  &  $\beta$ Pic \\
J22160+546  &  G 232-062  &  IC 2391/Argus  \\
J22509+499 &  2MASS J22505505+4959132  &  Columba \\
 
 \noalign{\smallskip}
\hline
\noalign{\smallskip}
\multicolumn{3}{c}{ \includegraphics[width=2mm]{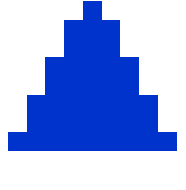} Age $\approx$ 150\,Ma} \\
\noalign{\smallskip}  
  
J02088+494*  &  G 173-039  &  AB Dor   \\
J02207+029  &  G 73-45  &  AB Dor  \\
J03332+462*  &  BD+45 784  &  AB Dor  \\ 
J03473$-$019*  &  G 80-021  &  AB Dor   \\ 
J04173+088*  &  LTT 11392  &  AB Dor   \\
J07163+331  &  GJ 1096 & AB Dor \\
J10289+008* & BD+01 2447 &  AB Dor \\
J14142$-$153  &  LTT 5581  &  AB Dor  \\ 
J23060+639*  &  MCC 858  &  AB Dor \\ 

  \noalign{\smallskip}
  \hline
  \noalign{\smallskip}

\multicolumn{3}{c}{\includegraphics[width=2mm]{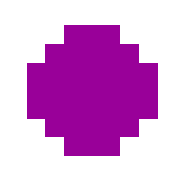} Age $\approx$ 300--800\,Ma} \\
  \noalign{\smallskip}

J00548+275*  &  G 69-32  &  Hyades  \\ 
J01214+243*  &  Ross 788  &  Hyades  \\  
J02256+375*  &  G 74-025  &  Hyades  \\ 
J03224+271*  &  LP 355-064  &  Hyades \\ 
J03548+163*  &  LP 413-108  &  Hyades \\
J03565+319*  &  HAT 214-02089  &  Hyades  \\ 
J04148+277*   &  HG 8-1 &  Hyades   \\
J04224+036*  &  RX J0422.4+0337  &  Hyades \\
J04227+205*  &  LP 415-030  &  Hyades  \\
J04238+149*  &  IN Tau  &  Hyades  \\
J04252+080N*  &  HG 7-207  &  Hyades  \\ 
J12156+526  &  StKM 2-809  &  Ursa Major  \\ 
J12417+567*  &  RX J1241.7+5645  &  Ursa Major  \\ 
J12485+495  &  RX J1248.5+4933  &  Ursa Major  \\
 
\noalign{\smallskip}
\hline
\end{tabular}
\tablefoot{
  \tablefoottext{a}{Asterisk (``*'') stands for confirmed members supported by the literature.}
}
\end{table}

To the 1508 FGKM stars in Praesepe and Hyades, we added bona fide young stars in our sample of 1581 stars (Sect.~\ref{sec.rv}) based on their kinematics and data from the literature. 
For that purpose, we selected 51 stars that have the same SKG classification in at least two of the three codes presented above. 
Of these, we kept 32 stars with unambiguous youth classification from the literature \citep[e.g.][]{Reid95_Hyades,ZS04,Binks14} and 11 with activity and rotation properties consistent with youth.
To this group, we then added the known Taurus-Auriga association member XEST~16-045 / J04206+272 \citep{Scelsi07}, which resulted in a list of 44 reference young stars, shown in Table~\ref{tab.55feten}. 

The aims of this analysis were to complement the Praesepe and Hyades stars with younger and cooler M dwarfs, to calibrate the dependencies of rotation and activity levels on age using the data of our CARMENES catalogue. We also aimed to assess whether the measured values strongly depend on age in the regime under $\sim$300\,Ma.
To that end, we separated stars with ages under 50\,Ma (i.e. sources belonging to Argus, IC~2391, $\beta$~Pictoris, Columba, Taurus, and Tucana-Horologium), between 50\,Ma and 300\,Ma (AB Doradus), and between 300\,Ma and 800\,Ma (Ursa Major, Hyades).
At this stage, we did not differentiate between Hyades cluster and supercluster because of their blurry boundary (e.g. \citealt{Meingast19,Roser19_hyades,Oh20,Jerabkova21,brandner23}).
Furthermore, eight of the Hyades stars in Table~\ref{tab.55feten} were also tabulated by \citet{Dou14}.

\begin{figure*}[!ht]
\centering
\includegraphics[width=0.45\hsize]{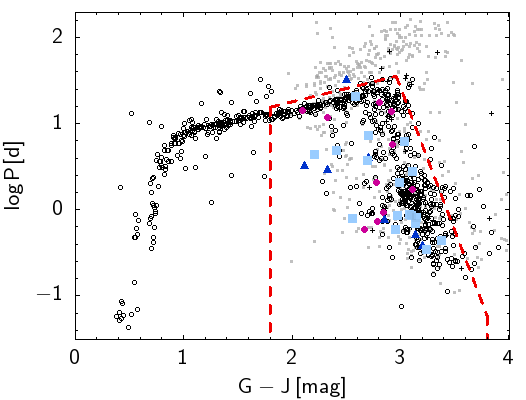}
\includegraphics[width=0.45\hsize]{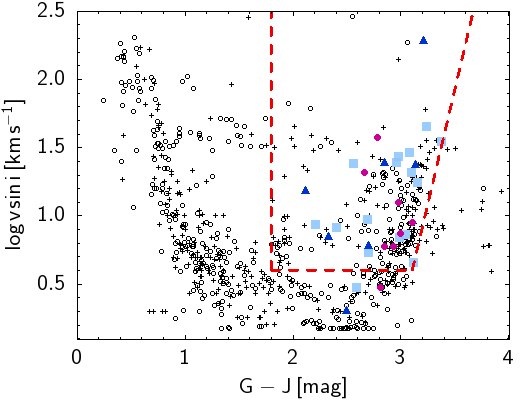}
\includegraphics[width=0.45\hsize]{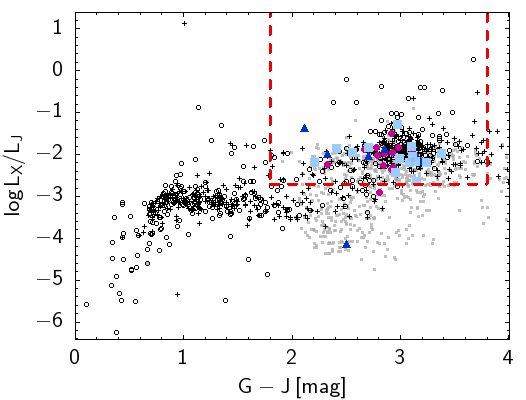}
\includegraphics[width=0.45\hsize]{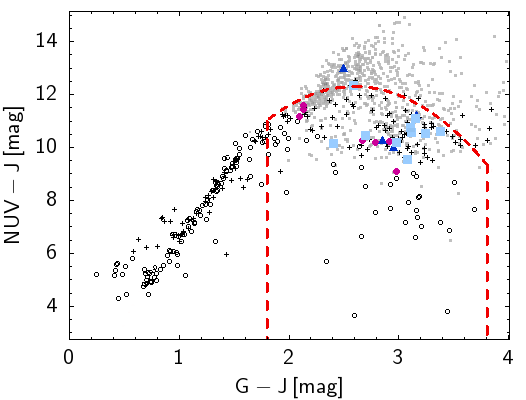}
\includegraphics[width=0.45\hsize]{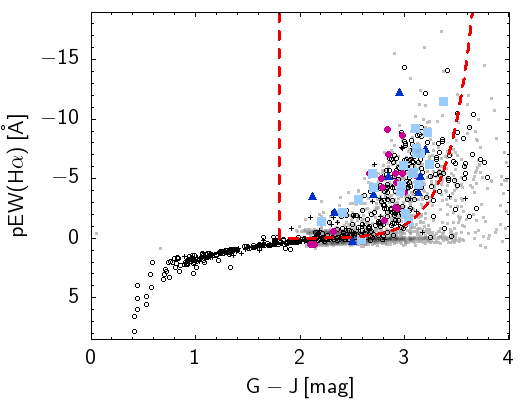}
\includegraphics[width=0.45\hsize]{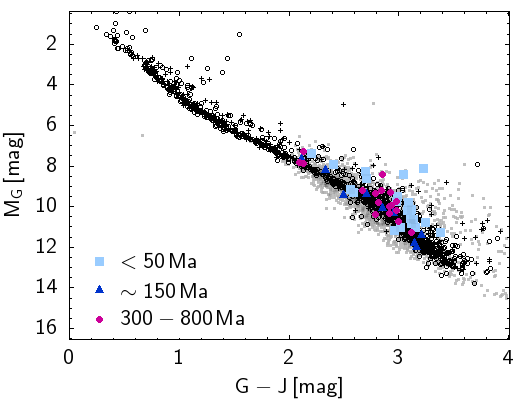}
\caption{ $P$, $v\sin{i}$, ${L}_{\rm X}/{L}_J$, $NUV-J$, pEW(H$\alpha$), and $M_{\rm G}$ as a function of $G-J$ colour, shown from left to right and top to bottom. Bona fide sources younger than 50\,Ma, of $\sim$150\,Ma, and between 300\,Ma and 800\,Ma are represented with light blue solid squares (\includegraphics[width=2mm]{Figures2023/sq_age50.png}), dark blue solid up triangles (\includegraphics[width=2mm]{Figures2023/tri_age150.png}), and purple solid circles (\includegraphics[width=2mm]{Figures2023/circle_age300.png}), respectively. 
Black open circles ($\circ$) and pluses ($+$) stand for Praesepe and Hyades cluster members, respectively, while grey dots  (\tgray{$\bullet$}) represent the 1581 Carmencita M dwarfs without known close companions at $\rho <$ 5\,arcmin.
The red dashed lines indicate the selection criteria from Eqs.~\ref{eq.prot}--\ref{eq.pew}.}
\label{fig.activity_limits}
\end{figure*}

Figure~\ref{fig.activity_limits} shows the rotational period, rotational velocity, X-ray emission ($\log{{\rm L}_X/{\rm L}_J}$), UV emission ($NUV-J$), and H$\alpha$ pseudo-equivalent width of our selected sample of young stars as a function of the $G-J$ colour, which is a proxy of the spectral type and, therefore, mass. 
The reason behind using the {\em Gaia}-2MASS colour is to avoid the impact of inaccurate spectral types in the sample. 
In Fig.~\ref{fig.activity_limits}, we also display the $M_G$ absolute magnitude-$G-J$ diagram, although we did not use it for any selection criterion. 
Stars are represented with different size symbols and colours, which designate ages under 50\,Ma, near 150\,Ma (AB~Doradus age), between 300 and 800\,Ma, as well as those of the Praesepe and Hyades stars. As a reference, we display in the background the 1581 stars of this study.

We found that all Praesepe, Hyades, and bona fide young stars consistently overlap in the $P$, $v\sin{i}$, X-ray, $NUV-J$ and pEW(H$\alpha$) diagrams.
From their positions in these diagrams, we distinguished two separate slopes that break at $G-J \sim$ 3\,mag (M3.0--4.0\,V) for $P$ and $v\sin{i}$, a parabola for $NUV-J$, an exponential for pEW(H$\alpha$), and a flat line for X rays.
In the colour-magnitude diagram, the position of the selected young stars from Table~\ref{tab.55feten} appears over or slightly above the main sequence, in line with the typical location of young stars \citep{Baraffe98,Cif20,Binks22}.

In light of these diagrams, we ended by defining five upper or lower limits for young active stars (up to 800\,Ma) that encompass over 85\% of Praesepe or Hyades stars and at least 90\% of the CARMENES bona fide stars in the $G-J$ = 1.8--3.8\,mag ($\sim$ K7--M6) range. These relations are as follows:

\begin{equation}
\begin{aligned}
 \log{P} &\leq 0.31\,{(G-J)}+0.63,\quad{ 1.8 \leq G-J \leq 3.0,}\\
 \log{P} &\leq -3.30\,{(G-J)}+11.30,\quad{3.0 < G-J \leq 3.8;}
 \end{aligned}
 \label{eq.prot}
\end{equation}

\begin{equation}
\begin{aligned}
\log{v\sin{i}} &> 0.602,\quad{ 1.8 \leq G-J \leq 3.1,}\\
\log{v\sin{i}} &\geq 1.04\,{(G-J)}-2.61,\quad{ 3.1 < G-J \leq 3.8;}
\end{aligned}
\end{equation}

\begin{equation}
\begin{aligned}
\log{{L}_{\rm X}/{L}_J} &\geq -2.7,\quad{ 1.8 \leq G-J \leq 3.8;}\\
\end{aligned}
\label{eq.xrays}
\end{equation}

\begin{equation}
\begin{aligned}
&NUV-J \leq -1.43+10.59\,{(G-J)}-2.04\,{(G-J)^2},\\
&\quad{ 1.8 \leq G-J \leq 3.8;}
\end{aligned}
\label{eq.nuv}
\end{equation}

\begin{equation}
{\rm pEW(H}\alpha) \leq -0.1\exp{(4.4\,{(G-J)-10.8})},\quad{ 1.8 \leq G-J \leq 3.8.}
\label{eq.pew}
\end{equation}

\noindent Here, the units are mag for $G-J$ and $NUV-J$, d for $P$, km\,s$^{-1}$ for $v \sin{i}$, {\AA} for pEW(H$\alpha$), and ${L}_{\rm X}/{L}_J$ is dimensionless. The relations are valid between $G-J = 1.8$\,mag and $G-J = 3.8$\,mag.
The selection based on $v\sin{i}$ may be incomplete since pole-on active stars with small inclination angles can be missed.
In column 6 of Table~\ref{table.name_pop_skg}, we provide for each star under analysis in the 1.8\,mag $< G-J <$ 3.8\,mag interval, an activity flag for each of the five equations presented here. Then,
`A' and `N' describe whether the star satisfies or not its corresponding relation (while no measurement available is represented by `.').

\section{Results and discussion}\label{sec.discussion}

\subsection{Activity-colour relations}

\begin{figure}
\centering
\includegraphics[width=\hsize]{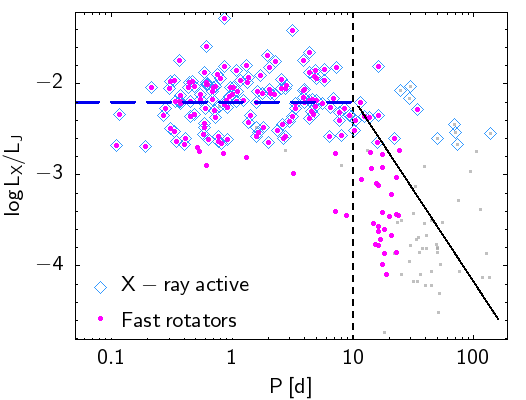}
\includegraphics[width=\hsize]{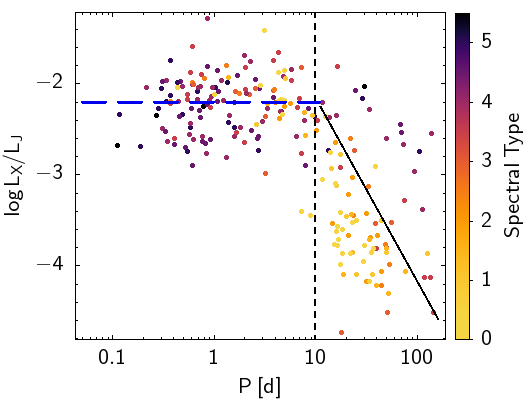}
\includegraphics[width=\hsize]{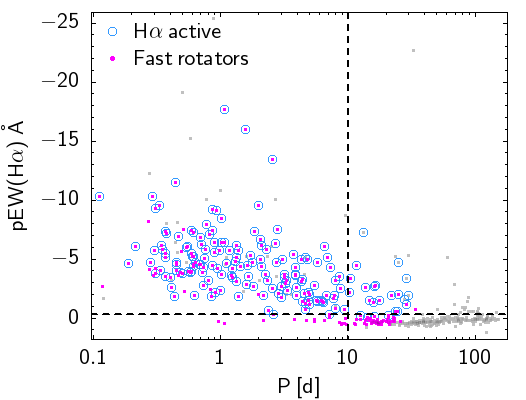}
\caption{X-rays emission (top and middle) and H$\alpha$ pseudoequivalent width (bottom) vs rotational periods of the Carmencita M dwarfs. 
Magenta filled circles represent fast rotating stars that fulfill Eq.~\ref{eq.prot}, blue open diamonds represent X-ray active stars that satisfy Eq.~\ref{eq.xrays}, and blue open circles represent chromospherically active stars that fulfill Eq.~\ref{eq.pew}. The complete sample of study is represented with grey dots.
Horizontal blue long-dashed lines in the top and middle panels represent the median saturation limit, while the black solid line follows a power law ${L_X/L_J} \propto P^2$.
Vertical and horizontal black dashed lines in the bottom panel mark the 10\,d and $-$0.30\,\AA\, boundaries of \cite{Shan24} and \cite{Schf19}, respectively.}
\label{fig.pewP}
\end{figure}

Several studies have aimed to describe the behaviour of activity and rotation with the spectral type.
Perhaps one of the most precise relation is that of rotation with spectral type observed in the top-left image of Fig.~\ref{fig.activity_limits}, where hundreds of accurate rotational periods of Praesepe and Hyades members depict the rotational slowdown of stars towards later spectral types. This behaviour has been widely studied to determine age-rotation relationships \citep[e.g.][]{Barnes10,Brown14,Dou19,Rebull17Praesepe,Jeff18,Spada20,Getman23}.

Coronal emission and rotation are closely connected, with rapid rotators showing large X-rays emission until a saturation level is reached \citep{Wright11,Reiners14}. This relation is also dependent on mass (i.e. on spectral type). In the top panel of Fig.~\ref{fig.pewP}, we display the X-ray emission with respect to rotational period.
We observe two slopes: one for fast-rotating stars in the X-ray saturated regime and another for slow-rotating stars that exhibit a steep decline in X-ray emission. The transition point occurs near 10\,d, consistent with findings from other researchers \citep{Magaudda20,Shan24}.
We highlight stars that are fast rotators or have active coronae, according to Eqs.~\ref{eq.prot} or \ref{eq.xrays}, respectively. Stars defined as fast rotators in this work have periods shorter than 34\,d (under 18\,d for 90\% of them). This second period decreases to 8\,d when considering stars that satisfy both relations simultaneously.
Stars with periods under $\sim$10\,d exhibit a median X-ray emission saturation limit of $\log{L_X/L_J} = -2.2 \pm 0.2$. To compare it with published values, we used of Eqs.~\ref{eq.lxljlbol} and \ref{eq.mbol} together with bolometric luminosities from \citet{Cif20}, and translated it into $L_X/L_{\rm bol} \sim -3.14$. This value is in agreement with other works (e.g. \citealt{Wright11,Reiners14}).
For comparing the turn-off point and slope of non-saturated stars, we computed the Rossby number that relates the stellar rotation period with convective flows:

\begin{equation}
    Ro=\frac{P}{\tau},
    \label{eq.rossby}
\end{equation}

\noindent{where $\tau$ is the convective turnover time, $\tau = 12.3\cdot \log{L_{\rm bol}/L_\odot}^{-1/2}$\,d when $P$ is in days \citep{Reiners2022}.
According to \citet{Wright11}, the saturation begins at $Ro \sim 0.13$, having stars with larger Rossby numbers longer periods and lower X-ray emission. 
However, $Ro \sim 0.13$ is a small value in light of our data. 
We obtained a value of $Ro=0.19$ after selecting stars with $\log{L_X/L_J} < -2.8$ (in the saturation level within $3 \sigma$) and periods over 8\,d.
In the non-saturated region,  we represent for comparison a power law ${L_X/L_J} \propto P^2$ as \citet{Reiners14}. From our plot, stars in this region seem to mostly follow a steeper slope. However, the large dispersion in our data prevented us from performing a solid correlation (however, see below).
We kept this analysis through $P$ rather than $Ro$ because of its direct comparison with observables and also because of the need of a reliable determination of the Rossby number in convection zones (see \citealt{Reiners14}).
The presence of stars with periods over 8\,d and below the X-ray saturation limit in the figure, could be explained by the age (as old as 800\,Ma) of the clusters used to define the inequalities. This age is sufficient for stars to have already undergone spin-down.

The same figure representing $L_X/L_J$ versus $P$ with a colour code indicating the spectral type of the stars is shown in the middle panel. The transition point at which the unsaturated regime starts appears to occur at longer periods with increasing spectral types (i.e. decreasing masses) \citep{Pizzolato03,KS07}. This is due to the slower rotational evolution of late type M dwarfs compared to early types \citep{New17,Magaudda20}. Also, the slope in ${L_X/L_J} \propto P^\beta$ seems to be spectral-type dependent, being steeper for early type stars (M0--2\,V) compared to mid-late type (up to M5\,V).

As for the $NUV-J$ index, a smooth parabola fit to our data better than a two-slope pattern (middle right panel of Fig.~\ref{fig.activity_limits}). We observe that UV emission steadily increases from early type M dwarfs towards mid type M dwarfs ($G-J = 2.6$\,mag, $\sim$ M3), where it reaches its maximum at $NUV-J = 12.3$\,mag, and gradually drops off towards later spectral subtypes. Hence, the $NUV-J$ colour varies only within two to three magnitudes for all spectral types in the M dwarf regime. This behaviour is in agreement with what \cite{Schneider18} observed from the flux ratios, $F_{NUV}/F_J$ and $F_{FUV}/F_J$, together with a noticeable decay with age for early type M dwarfs compared to later types (see Fig.~6 therein).

In the H$\alpha$ diagram on the left bottom corner of Fig.~\ref{fig.activity_limits}, an exponential increase towards later spectral subtypes offers a better fit than a flat trend, as established by other authors \citep[e.g.][]{West15,New17,Schf19}.
This exponential pattern resembles the empirical criteria to classify T~Tauri stars and substellar analogues, but shifted to lower pEW(H$\alpha$) values \citep{Barrado03,WB03,Fang10}. 

We represent pEW(H$\alpha$) versus $P$ on the bottom panel in Fig.~\ref{fig.pewP} and highlight stars that are fast rotators or chromospherically active according to Eqs.~\ref{eq.prot} or \ref{eq.pew}, respectively. 
Our magnetically active stars have pEW(H$\alpha$) $\leq -$0.3\,\AA\ as in \cite{Schf19} in all but four cases that have -0.25<pEW(H$\alpha$)<-0.05\,\\A and periods between 12 and 21 days.

\subsection{Kinematic classification}

As described in Sect.~\ref{sec.pop}, we classified the 2187 sources in Carmencita with radial velocities, distances, and proper motions into young (YD), thin (D), thick-thin transition (TD-D) and thick (TD) disc members, and halo (H) population stars.
The population assignation of each star in the catalogue is listed in column 2 of Table~\ref{table.name_pop_skg}, while the number of targets ascribed to each population and corresponding ratios are presented in Table~\ref{table.pop}. The Toomre and B\"ottlinger diagrams in Fig.~\ref{fig.boettlinger+toomre} show the candidate members for each population.
The latter diagrams illustrate Eggen's boundaries for young stars in the $(U,V)$ and $(V,W)$ planes, along with the location of sources belonging to the different SKGs assigned by {\tt SteParKin} (see below).
As expected \citep{Chen01,Juric08}, the majority of the stars (1958) belong to either the thin disc (1245) or the young disc (713) of the Galaxy. 
On the other hand, only 78 stars belong to the thick-thin transition disc and 149 to the thick disc.
The remaining two stars are halo candidates: Ross~53A / J14575+313 and LP~651-7 / J02462$-$049. 
Both of them are high proper-motion field stars ($\mu = 2.5$\,arcsec/a and $\mu = 1.4$\,arcsec/a, respectively) and had indeed been classified as halo members by \cite{greensteinhalo}.

\begin{figure*}
\includegraphics[width=0.49\hsize]{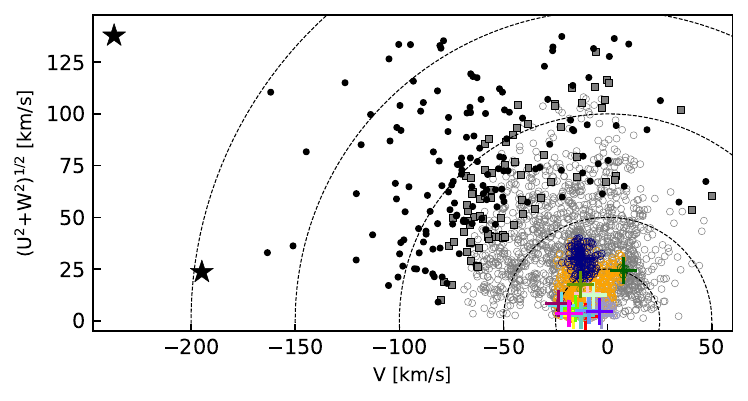}
\includegraphics[width=0.49\hsize]{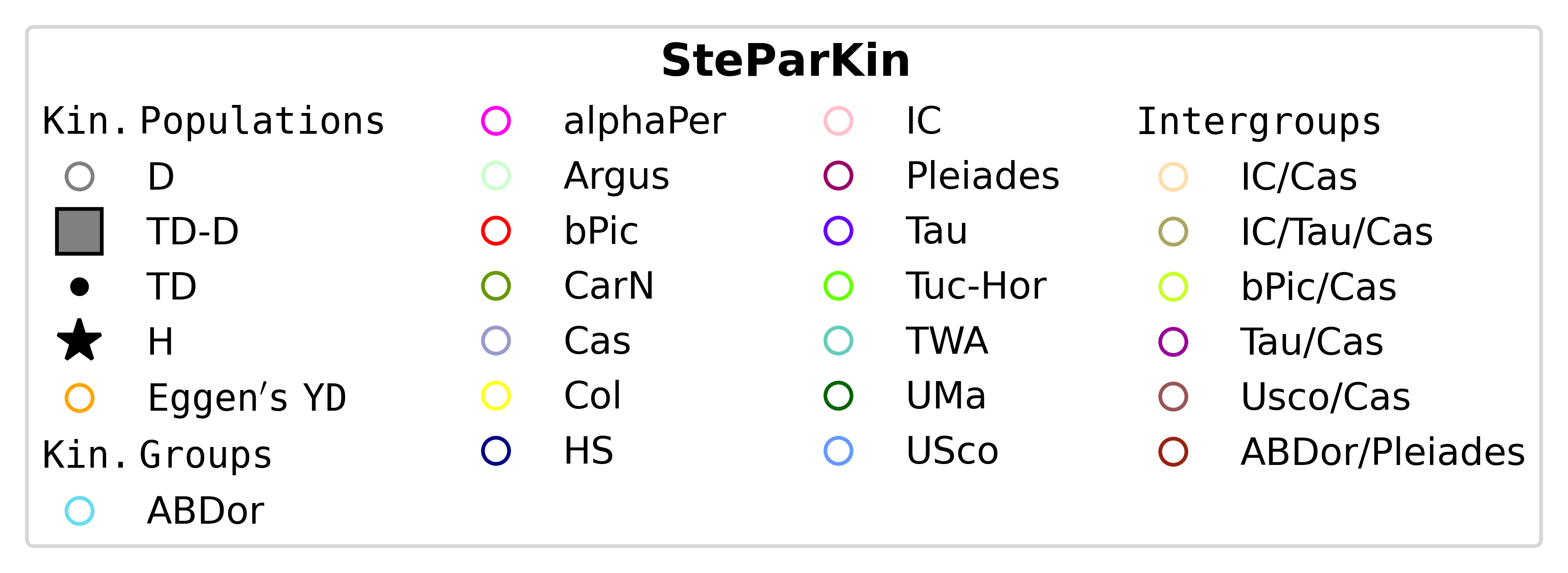}
\includegraphics[width=0.49\hsize, trim=0 0 12cm 11cm, clip]{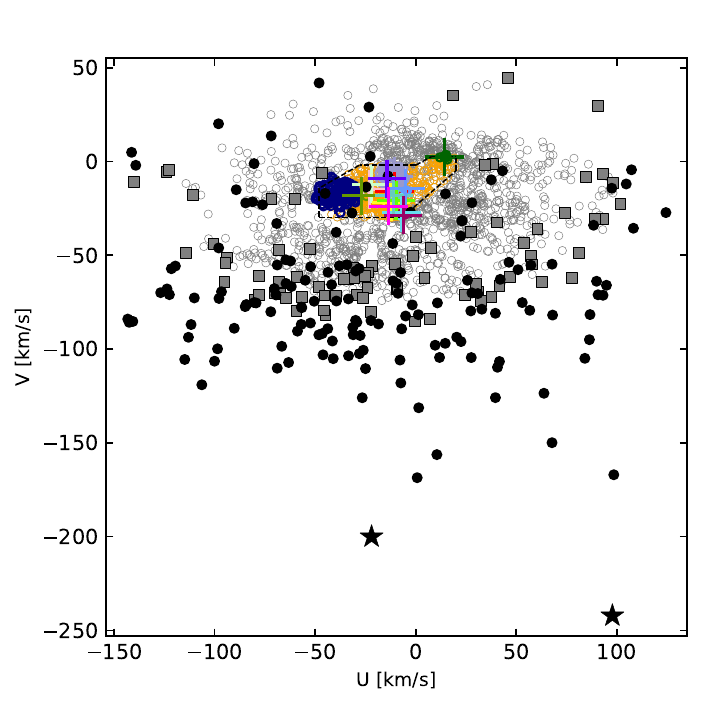}
\includegraphics[width=0.49\hsize, trim=12cm 0 0 11cm, clip]{Figures2023/SteParKin_UV_WV.pdf}
\caption{Toomre ($({U}^2+{W}^2)^{1/2}$ vs $V$, \textit{top-left}) and B\"ottlinger ($V$ vs $U,$  \textit{bottom-left}, and $V$ vs $W$, \textit{bottom-right}) diagrams.
Filled star symbols stand for candidate stars in the halo, filled black circles for the thick disc, grey filled squares for thick-thin transition disc, and open circles for the thin disc. 
Colour-coded open circles are stars in young SKGs as by {\tt SteParKin} (legend shown in the \textit{top-right}).
Filled crosses show the center of each group. 
In the Toomre diagram, the dashed lines indicate constant space velocities with respect to the local standard of rest, $v=(U_{\rm LSR}^2+V_{\rm LSR}^2+W_{\rm LSR}^2)^{1/2}$, at 25, 50, 100, 150, and 200\,km\,s$^{-1}$.
In the B\"ottlinger diagrams, the dashed lines indicate Eggen's young disc population criteria.}
\label{fig.boettlinger+toomre}
\end{figure*}

\begin{table}
\centering
\caption{Number of stars per stellar population and per sample.}
\label{table.pop}
\begin{tabular}{l cc c cc}
 \noalign{\smallskip}
 \hline
 \hline
 \noalign{\smallskip}
Stellar population       &  \multicolumn{2}{c}{All RV stars} & \multicolumn{2}{c}{Single$^a$ RV stars}\\
       &   Num. & Ratio [\%] & Num.  &  Ratio [\%]\\
\noalign{\smallskip}
 \hline
\noalign{\smallskip}
Young disc & 713 &32.6  &  480 & 30.4   \\
Thin disc &     1245 &56.9  & 922 & 58.3  \\
Thick-thin disc &       78 &3.6 &  67    &4.2   \\
Thick disc & 149 &6.8 &  111 &  7.0  \\
Halo & 2 & 0.1 & 1 & 0.06 \\
\noalign{\smallskip}
\hline
\end{tabular}
\tablefoot{
\tablefoottext{a}{``Single'' refers to stars without known close companions.}
}
\end{table}

\begin{figure}
\centering
\includegraphics[width=\hsize]{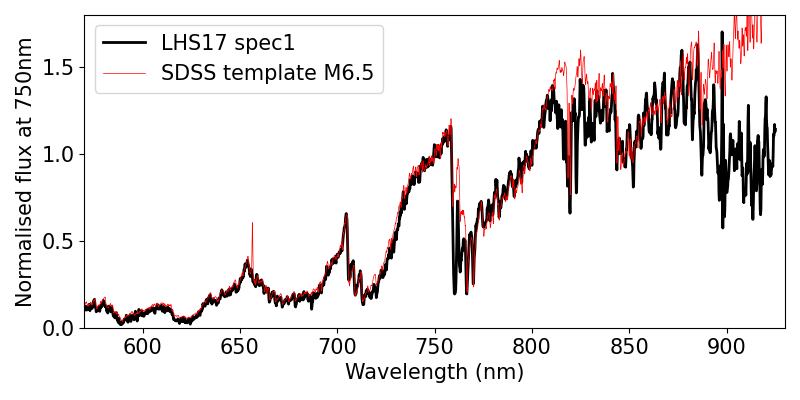}
\includegraphics[width=\hsize]{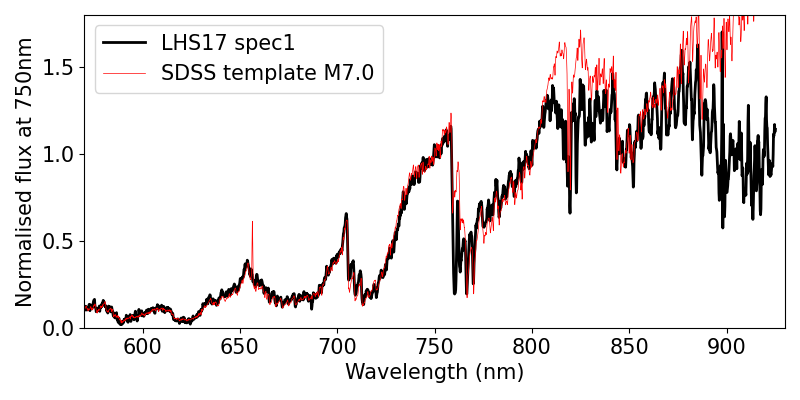}
\caption{Comparison of ALFOSC spectra of LP 651-7 with Sloan templates.}
\label{fig.spec}
\end{figure}

Ross~53A is a subsolar-metallicity M2 dwarf with a companion at 0.8\,arcsec, whose close multiplicity prevented us from including it in our activity-rotation analysis \citep{Ross25,Stassun19,WDS,Birky20}.
The other halo candidate star, LP~651-7, has not been investigated in depth \citep{Winters15,New17}.
Previously, \citet{Reid95} had classified it as a regular M6 dwarf.
We secured a low-resolution optical spectrum of LP~651-7 with the Alhambra Faint Object Spectrograph and Camera (ALFOSC) on the 2.5\,m Nordic Optical Telescope on a service night on 15 November 2022. 
We exposed for 600\,s with the grism \#20 and the slit of 0.75\,arcsec, which yielded a spectral resolution of approximately 1000 at central wavelength 7850\,{\AA}.
We carried out the standard data reduction under the IRAF environment \citep{Tody86,Tody93} and wavelength-calibrated the extracted spectrum with He/Ne/Th-Ar arc lamps collected on the same night.
After instrumental-response calibration with a spectro-photometric standard, we compared our spectrum to other of slightly higher spectral resolution (1.79\,\AA) but narrower wavelength coverage and a lower signal-to-noise ratio kindly provided to us by K.\,L.~Cruz.
We classified LP~651-7 as a M6.5--M7.0 field dwarf by direct comparison with Sloan templates \citep{Bochanski07} in Fig.~\ref{fig.spec}, consistent within uncertainties with the classification of \citet{Reid95}.
We did not observe any sign of lack of metals in any of the low-resolution spectra of LP~651-7 available to us, suggesting a metallicity close to solar.
In addition, we compared its tangential velocity, $v_{\rm tan}$, with that of stars in the different populations from this work (see Fig.~\ref{fig.vtan}). While having the second highest $v_{\rm tan}$ in our sample (after Ross~53A) of 200.4\,km\,s$^{-1}$, this value is still in agreement with the tangential velocities displayed by thick-disc stars in our sample. It also falls very close both in terms of $v_{\rm tan}$ and distance to other thick-disc members like Ross~769 / J21048$-$169, with $v_{\rm tan} \approx$ 194.1\,km\,s$^{-1}$. Although we do not claim any incompatibility with a halo star based on the value of its tangential velocity, this fact (coupled  with the lack of metal content in the spectra and its position in the diagrams of Fig.~\ref{fig.boettlinger+toomre}, i.e. close to the thick-disc population) led us to propose the star as a thick-disc M dwarf, rather than a halo member.
As a result, there is only one halo member in the Carmencita sample: Ross 53A / J14575+313.

In Fig.~\ref{fig.vtan}, we display tangential velocities as a function of the distances of all the 2187 stars with radial velocities. As a reference, we use the work of \citet{Zhang18}, who presented a sample of halo and thick-disc low-mass sub-dwarfs with tangential velocities well above 100\,km\,s$^{-1}$. In contrast, the median $v_{\rm tan}$ of a defined sample of 5000 field main-sequence stars (mostly with spectral types of FGK and M0--4) in the thin disc was 36\,km\,s$^{-1}$. Hence, field stars display low $v_{\rm tan}$ compared to older populations.
In our sample, mean values of the tangential velocity of stars in the young and thin disc populations (24 and 40\,km\,s$^{-1}$, respectively) are in agreement with the 36\,km\,s$^{-1}$ median value of \cite{Zhang18}. 
Mean values of $v_{\rm tan}$ increase to 75 and 88\,km\,s$^{-1}$ for thick-to-thin transition disc and thick-disc stars, respectively. The one unambiguous halo member exhibits $v_{\rm tan} = 272$\,km\,s$^{-1}$.
The similarity between $v_{\rm tan}$ of thick and thick-to-thin transition discs results from their partial overlapping.

Finally, Table~\ref{table.name_pop_skg} lists, for each star, the SKG assigned by {\tt SteParKin}, {\tt LACEwING}, and {\tt BANYAN} $\Sigma$, together with their probabilities (when available). 
We imposed a minimum membership probability of 20\% in both {\tt LACEwING} and {\tt BANYAN} $\Sigma$.
This kinematic classification will be employed in the youth assessment described below.

\subsection{Galactic populations and activity}

 In this section, we discuss how the 1581 stars without known close companions ($\rho <$ 5\,arcsec; Sect.~\ref{sec.rv}) behave in terms of rotation and activity in each galactic population. We also make a comparison with previous results of related works in the literature.
On average, about 28\% of the original 2187 stars are close multiples according to our definition, although the fractions vary from 18.8\% in thick-thin stars to 30.6\% in young-disc stars.
We summarise the number of stars without close companions belonging to each population in Table~\ref{table.pop}.

The rotational period has been widely used as a proxy for age, as young stars tend to rotate faster than old stars \citep{Bouvier93,KS07,Medina22,Shan24}. 
As a comparison, \cite{New16} presented an analysis of the rotation-age relation in a stellar sample that resembles ours. 
They found a gap in the period-spectral type diagram that increases with decreasing mass (i.e. towards late spectral types, see Sect.~8). In particular, they observed that stars with intermediate (10--70\,d) periods are uncommon.
This gap might also appear in the top left panel of Fig.~\ref{fig.activity_limits} between the sequence defined by Praesepe and Hyades cluster members and that of our field M dwarfs.
These authors also concluded that stars with $P < 10$\,d are younger than 2\,Ga on average, while stars with $P > 70$\,d are older than 5\,Ga. 
Although our data do not allow us to determine precise ages, we assessed typical periods of stars in the different galactic populations. 
In our sample, 61\% of stars with $P < 10$\,d belong to the young disc, whereas 78\% of the stars with $P > 70$\,d belong to the rest of older populations.
We computed the median values of the rotational period at 5.5, 34.4, 61.3, and 46.6\,d for stars in the young, thin, thick-thin, and thick discs, respectively. 
The median $P$ for the stars with at least one SKG classification by {\tt SteParKin}, {\tt LACEwING}, or {\tt BANYAN} $\Sigma$ is even shorter, namely 3.3\,d.
Similarly, the median $v\sin{i}$ for stars in the young, thin, thick-thin, and thick discs are 4.0, 3.0, 2.5, 2.5\,km\,s$^{-1}$, respectively, and 7.4\,km\,s$^{-1}$ for stars in SKGs.
As a result, there is indeed an age trend from the youngest stars in SKGs to the oldest stars in the thick-thin and thick discs in our rotation data (see also Fig.~11 in \citealt{Shan24}).

\begin{figure}
\centering
\includegraphics[width=\hsize]{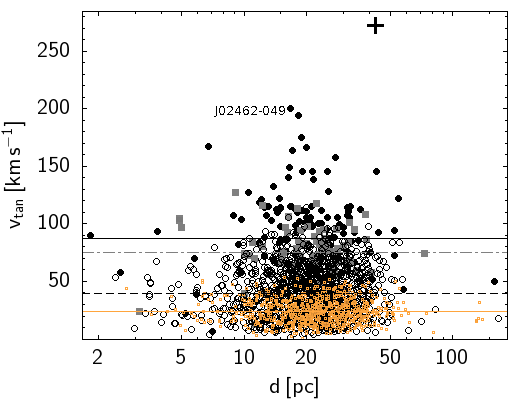}
\caption{Tangential velocity vs distance for the 2187 stars in the Carmencita sample with radial velocities. Each galactic population is represented with colours and symbols as in Fig.~\ref{fig.boettlinger+toomre}. Black solid, grey dash-dotted, black dashed and amber solid lines represent the mean values of $v_{\rm tan}$ of the thick, thick-thin, thin, and young discs populations. LP~651-7 / J02462$-$049 is the former halo candidate.}
\label{fig.vtan}
\end{figure}

\cite{West15} also studied the chromospheric activity-rotation connection from H$\alpha$ emission, and observed that M1--4\,V stars with periods $P < 26$\,d and M5--8\,V stars with periods $P < 86$\,d are all magnetically active according to their own definition (pEW(H$\alpha$) $<$ $-$0.75\,\AA). We were able to compare their flat activity criterium with the exponential one of this work.
In our sample, while imposing the same constraints, only 76\% and 32\% of the corresponding stars are chromospherically active using Eq.~\ref{eq.pew}.
To assess such a dramatic difference, we went back to our subsample of H$\alpha$ active stars from Eq.~\ref{eq.pew} and with rotation periods.
We found that 90\% of the chromospherically active M1--4\,V (actually M1.0--4.5\,V) stars have $P < 14.6$\,d (and 99\% have $P < 28.5$\,d), which is consistent with the results of \cite{West15}.
However, 90\% of the chromospherically active M5--8\,V stars have $P < 7.2$\,d.
On the one side, we noted the poor statistics, as relatively few late-type stars have precise $P$ determinations.
For example,  in our sample, there are 74 stars with spectral type M5.0\,V or later and a $P$ determination, with only 18 of them being H$\alpha$ active (according to Eq.~\ref{eq.pew}).
Furthermore, a large proportion of such $P$ values for late dwarfs have been reported relatively recently with data from either the ground \citep{DA19,Donati23,Fou23,Fuhr23}, space \citep{Rae20,Magaudda22}, or both \citep{Shan24}.
On the other hand, we ascribe most of the difference at the latest spectral types to our different definition of H$\alpha$ activity (flat by \citealt{West15}, exponential by us).

Similarly to our approach to $P$ and $v\sin{i}$, we investigated the average pEW(H$\alpha$) of stars in different galactic populations.
However, since it strongly varies as a function of spectral type, we used the 90\% percentile (Q90) instead of the median.
The Q90 of our pEW(H$\alpha$) are $-$6.0, $-$3.0, $-$0.2, $-$0.6\,{\AA} for stars in the young, thin, thick-thin, and thick discs, respectively, and $-$7.1\,{\AA} for stars in SKGs.
Although the spread is large, this result is again in line with previous works \citep[e.g.][]{West06,Kiman21}.

Finally, \cite{Jon16} showed that NUV emission decreases with distance from the galactic plane (i.e. with age) for mid-M dwarfs. 
We confirm this behaviour with a new observable, which is the variation of the ratio of $NUV-J$ active stars for each Galactic population:
up to 34.6\%, 15.1\%, and 6.5\% of the M dwarfs (regardless their spectral type) in the young, thin, and thick discs are active, according to Eq.~\ref{eq.nuv}, respectively (noting that there are no $NUV-J$ active stars in the transition disc or halo).
This ratio increases to 42.2\% when we consider only the stars with at least one SKG classification by {\tt SteParKin}, {\tt LACEwING}, or {\tt BANYAN} $\Sigma$.
We applied the same methodology to X-ray emission, and found that 56.6\%, 35.0\%, and 40.0\% of the M dwarfs in the young, thin, and thick discs are active according to Eq.~\ref{eq.xrays}, respectively (we included the only X-ray active transition-disc star in the thick-disc statistics).
The ratio of X-ray active stars in SKGs increases to 62.3\%. 
From the abnormally high ratio of X-ray active stars in the thick disc, we can conclude that X-rays are not as good as $NUV-J$ in helping assess the age of relatively old stars. 
This result is a consequence of the distinct saturated and non-saturated coronal X-ray emission regimes in M dwarfs, which has been extensively discussed in the literature \citep{Stauffer94,Pizzolato03,Wright18,Magaudda20,Reiners2022}.

\subsection{Young stars}\label{sec.activeandyoung}

\begin{figure}
\centering
\includegraphics[width=\hsize]{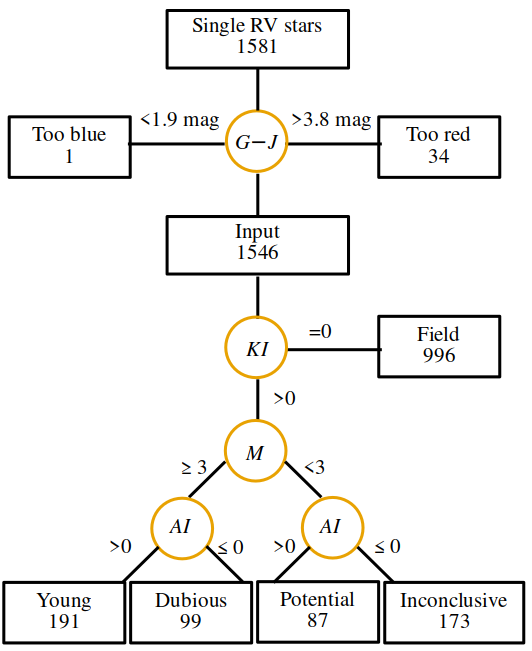}
\caption{Flowchart for the classification of young and active stars in our sample.}
\label{fig.diagramaflujo}
\end{figure}

Lithium is destroyed in M dwarfs in about 30\,Ma \citep{Chabrier2000, JacksonJeffries2014a, Galindo-Guil2022} and references therein).
Of the 382 M dwarfs with telluric-absorption-corrected, high S/N, optical and near-infrared template CARMENES spectra collected by \citet{Nagel23}, only two stars, namely AU~Mic and 2MASS J05082729--2101444, display the Li~{\sc i} $\lambda$6707.8\,{\AA} doublet in absorption. 
Very few of the other $\sim$1800 M dwarfs that have not been investigated with CARMENES have spectra of a high-enough resolution and S/N to look for Li~{\sc i}, which prevents any lithium-based age dating for virtually all the stars in our sample.
Both AU~Mic and 2MASS J05082729--2101444 are young stars in the $\beta$~Pictoris moving group that have been included in several works based on CARMENES \citep{Cale21,Mallorquin24,Kaur24}.

Instead of relying in the stellar lithium abundances, we took into account the accumulated kinematic and activity data on our stars without close companions for identifying the youngest ones.
For that, we defined two ad hoc indices, namely, the `kinematic index' ($KI$) and `activity index' ($AI$).
For computing $KI$ for each star, we counted the positive answers to the question whether it kinematically belongs to the young disc or a SKG according to {\tt SteParKin}, {\tt LACEwING}, or {\tt BANYAN} $\Sigma$. As a result, $KI$ can vary between 0 and +4.
The maximum $KI$ is for stars in the young disc with SKG assignation by the three codes, whether they are identical.
Likewise, for computing $AI$ for each star, we added 0.4 for each of the corresponding activity parameters satisfying Eqs.~\ref{eq.prot}--\ref{eq.pew} or subtracted 0.2 when it did not satisfy them (i.e. `A' or `N' in Table~\ref{table.name_pop_skg}, respectively). As a consequence, $AI$ can vary between --1 and +2.
The maximum $AI$ is for stars that satisfy the five Eqs.~\ref{eq.prot}--\ref{eq.pew} simultaneously. The $AI$ index is biased, as it does not only account for the strength of the different activity indicators, but also its measurement.
Naturally, surveys tend to collect data of the brightest objects and, therefore, the brightest stars tend to have the five measurements ($P$, $v\sin{i}$, $L_{\rm X}/L_J$, $NUV-J$, pEW(H$\alpha$)) simultaneously.

After computing the $KI$ and $AI$ indices and collecting the number of activity parameters with available measurements, $M$, we applied the decision tree summarised in the flow diagram in Fig.~\ref{fig.diagramaflujo}. We classified the stars into field, young, potentially young (dubbed 'potential'), dubious, and inconclusive.
Here, field stars have $KI = 0$, that is, they are not members of the galactic young disc nor belong to any SKGs according to {\tt SteParKin}, {\tt LACEwING}, or {\tt BANYAN} $\Sigma$.
Also, dubious stands for stars that have young kinematics ($KI>0$) and more than two data points for which their current activity-rotation data points towards non active stars ($AI<0$), although more data would be needed to refute them.
Potential stars have young kinematics ($KI>0$) as well and fewer than three data points. They have activity or rotation signatures ($AI>0$) but still additional measurements would be needed to confirm youth.
Inconclusive stands for stars again with young kinematics ($KI>0$) and without enough data to evaluate youth.
The remaining stars, which we call young, have at least one kinematic assignation to the young disc or a SKG, at least three measurements of activity indicators ($M \ge 3$), and an $AI$ greater than zero.
In Fig.~\ref{fig.counter}, we illustrate the results of our quantitative classification. The values +1 for $KI$ and +0.4/$-$0.2 for $AI$ were chosen for maximising the separation between active and non-active stars in such a diagram, while keeping the quantisation as simple as possible.
The subtraction for $AI$ was intended to better discriminate between stars with measurements available for evaluation that are not active, and stars without measurements available.

\begin{figure}
\centering
\includegraphics[width=\hsize]{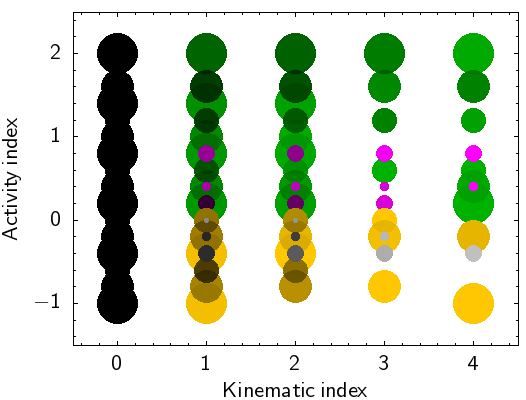}
\caption{Activity index, $AI$, vs kinematic index, $KI$. 
Young stars are green, potential are magenta, dubious are yellow, inconclusive are grey, and field are black.
Sizes of the circles are proportional to the number of measurements, $M$.}
\label{fig.counter}
\end{figure}

Of the 1581 RV stars without known companions, there is one star with a colour of $G-J <$ 1.9\,mag (TYC 4450-1440-1 / J20109+708, K5\,V) and 34 stars with a colour of $G-J >$ 3.8\,mag, which are either too blue (warm, early) or too red (cool, late) for the range of application of Eqs.~\ref{eq.prot}--\ref{eq.pew}.
This additional filter left us with 1546 stars for which we actually computed $KI$ and $AI$.
Of them, 996 are field stars with $KI = 0$ regardless of their activity, while of the remaining 550 stars with $KI > 0$, we identified 191 young, 87 potential, 99 dubious, and 173 inconclusive stars.
The fraction of young stars among the 1546 low mass stars is in the range between 12 and 24\%, considering also the 87+99 potential and dubious samples. There is another 11\% corresponding to the 173 inconclusive stars that may increase this fraction up to 35\%.

For the 191 young stars, we searched the literature for previous references to age determinations (e.g. \citealt{Shk09}), SKG membership (e.g. \citealt{Zuck11,Shk12,Schlieder12a,Malo13,Gagne15,AF15betapic}), and T~Tauri status (e.g. \citealt{Li2000}). We found information for 78 stars only, including the 33 stars marked with an asterisk in Table~\ref{tab.55feten} of bona fide members of young SKGs.
Our literature analysis is summarised in footnote `e' of Table~\ref{table.name_pop_skg}.
We were not able to identify any age-related reference to any of the remaining 113 stars.

Very close binaries may suffer from tidal locking, speeding up, and enhanced magnetic activity \citep{Holzwarth03,Meibom05,olah07,Kovari17}.
Of the 113 young star candidates, 36 were also identified by \cite{Cifuentes24} as new very close binary candidates, including spectroscopic binaries, based on \textit{Gaia} data (for this work only confirmed binaries were removed for the analysis).
Most (if not all) of these very close binary candidates may have $AI > 0$ because of actual binarity and subsequent tidal locking. 
We thus have 77 new validated young star candidates with no known close companions, which we mark with footnote `f' in Table~\ref{table.name_pop_skg}.
Similarly, 24 of the 87 potentially young stars have a very close companion candidate that may explain their excess of activity not necessarily related to youth.
Furthermore, some unreported close binaries in the field may have not been identified in previous multiplicity surveys, which may explain their moderate chromospheric or coronal activity (see Sect.~\ref{sec.the37}).

\begin{table}
\setlength{\tabcolsep}{2.0pt}
\centering
\caption{26 new young candidates  identified as highly active in this work.} 
\label{table.12singlenew}
\scriptsize
\begin{tabular}{l l l l c }
 \noalign{\smallskip}
 \hline
 \hline
 \noalign{\smallskip}
 Karmn & Name & Pop. & SKG & Activity   \\
        &       &       &&flag\tablefootmark{a}         \\
\noalign{\smallskip}
 \hline
\noalign{\smallskip}
J01033+623  &   V* V388 Cas & YD  & ... & AAA.A \\
J01134$-$229 &  GJ 1033 & YD &...&  A.AAA               \\
J02071+642 &  G 244-049  &  YD &  IC 2391 &  AAA.A\\
J03236+056  &   1RXS J032338.7+054117   & YD  & TW Hya / $\beta$ Pic    & A.AAA  \\
J03361+313 &  [GBM90] Per 49 &  YD &  Castor &  AAA.A\\
J04122+647  &   G 247$-$015 &  YD & Hyades   & AAA.A  \\
J04304+398  &   V546 Per &  YD & ...  & AAA.A  \\
J05566$-$103 &  1RXS J055641.0-101837  &  YD & ...  & AAA.A\\
J06000+027  &   G 99$-$049 &  YD & ...  & AAA.A  \\
J09161+018 &  RX J0916.1+0153 &  D  &  UMa &  AAA.A\\
J09302+265 &  2MASS J09301445+2630250 &  YD &  ... &  AAAAA\\
J09589+059  &   NLTT 23096  &  YD & Castor  & AAAAA  \\
J10384+485 &  LP 167-071 &  YD &  Hyades SC &  AAAAA\\
J10522+059 &  NLTT 25568 &  YD &  ... &  AAAAA\\
J13518+127 &  RX J1351.8+1247 &  D  &  UMa &  AAA.A\\
J14194+029  &   NLTT 36959  & YD  &  ... & A.AAA  \\
J14227+164  &   NLTT 37131  & YD  & Hyades  & AA.AA  \\
J14312+754  &   2MASS J14311348+7526423 &  YD & ...  & AAAAA  \\
J14321+160  &   LP 440$-$038    & D  & UMa  & AAAAA  \\
J15480+043 &  RX J1548.0+0421 &  YD &  ... &  AAA.A\\
J15557+686 &  RX J1555.7+6840 &  YD &  Castor &  A.AAA\\
J16401+007  &   LP 625$-$034    & YD  & UMa  & AAA.A  \\
J16570$-$043  &   LP 686$-$027    &  YD & ...  & AAAAA  \\
J18498-238  &   V1216 Sgr    & D  & Castor  & AA.AA  \\
J19422-207  &   2MASS J19421282$-$2045477    & YD  & ...  & AAA.A  \\
J19511+464 &  G 208-042  &  YD &  IC 2391/Argus  &  AAA.A\\

\hline
\end{tabular}
\tablefoot{
\tablefoottext{a}{In the five-character activity flag, `A' and `N' indicate whether the star satisfies each of the Eqs.~\ref{eq.prot}--\ref{eq.pew}. Non literature values for assessment are represented with `.'.
The five labels refer to P$_{\rm{rot}}$, $v\sin{i}$, $\rm{L}_X/\rm{L}_J$, $NUV-J$, and pEW(H$\alpha$) in this order.
}}
\end{table}

In Table~\ref{table.12singlenew}, we present the 26 new young star candidates with $AI \ge$ 1.5, namely, those with the largest number of activity indicators satisfying Eqs.~\ref{eq.prot}--\ref{eq.pew}.
We suggest a better characterisation and SKG assignation in forthcoming work for these 77 newly validated young star candidates.

\subsection{The most active stars}\label{sec.the37}

\begin{figure}
\centering
\includegraphics[width=\hsize]{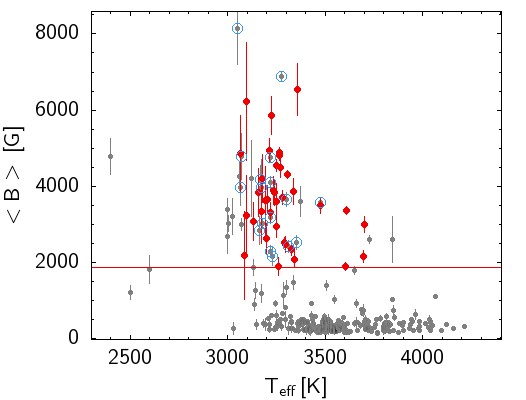}
\caption{Averaged magnetic field strength as a function of $T_{\rm eff}$.
The 36 very active stars of this work are plotted with red filled circles, 
the 17 stars with fewer activity features and strong magnetic fields are indicated by the blue empty circles, while the sample of M dwarfs in \cite{Reiners2022} is given by the grey circles.
The red solid line is the 1900\,G lower limit.}
\label{fig:magneticfields}
\end{figure}

Of the 1546 input M dwarfs (stars with colours 1.9\,mag $< G-J <$ 3.8\,mag), there are only 118 stars that satisfy at least four of the five activity criteria of Eqs.~\ref{eq.prot}--\ref{eq.pew}.
Of them, 81 stars belong to the young disc, 35 to the thin disc (including 16 young stars in SKGs), and 2 to the thick disc. Hence, 97 stars are identified as young based on our criteria and 9 of them are associated to SKGs under 50\,Ma; the 2 thick-disc stars are Ross~388 / J05091+154 and LP~86-173 / J06054+608.
Ross~388 shows a significant radial velocity variability from \textit{Gaia} DR3 data \citep{Katz2023}, which might be related to such high activity features. 
The presence of an unresolved binary companion \citep{Cifuentes24} may also explain such high activity features in a supposedly old star \citep{Shk09,Jeff18,RM2020_Romy,Rae20}.
This close multiplicity was not detected, however, in lucky imaging observations by \cite{Jodar13} and \cite{Cor17}.
We conducted a thorough search in the literature on the other thick star (i.e. LP~86-173) and found that it had incorrectly passed our initial close-binary filters. 
It was identified as a single-lined spectroscopic binary of a white dwarf primary and an M-dwarf companion \citep{Winters20}. 
The white dwarf has a similar mass, but it is hardly visible in photometry or low-resolution spectroscopy.
From this finding, we concluded that ($i$) the scenario of a white dwarf and an M dwarf in a close orbit would effectively explain its high activity (e.g. $P \approx$ 0.31\,d, $v\sin{i}$ = 27.9~$\pm$~1.3\,km\,s$^{-1}$, pEW(H$\alpha$) = $-$9.3~$\pm$~ 0.6\,{\AA}, $NUV - J$ = 8.509~$\pm$~0.034\,mag -- \citealt{AF15,New16,Kess18,Winters20,Cif20}),
($ii$) the existence of an evolved white dwarf in the system supports our kinematic thick-disc classification, 
($iii$) some other close spectroscopic binary may have passed our initial filters,
and ($iv$) this unnoticed close multiplicity may have also contaminated previous searches for young stars \citep{Shk11,Klutsch14,Gagne14BanyanII}.

\begin{table}
\setlength{\tabcolsep}{1.9pt}
\centering
\caption{36+17 stars with strong magnetic fields from \cite{Reiners2022}.} 
\label{table.27}
\scriptsize
\begin{tabular}{l l l l c c}
 \noalign{\smallskip}
 \hline
 \hline
 \noalign{\smallskip}
 Karmn\tablefootmark{a} & Name & Pop. & SKG & Activity & Class\tablefootmark{c}  \\
        &       &       &&flag\tablefootmark{b} &       \\
\noalign{\smallskip}
 \hline
\noalign{\smallskip}
\multicolumn{6}{c}{\includegraphics[width=3mm]{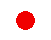} \em Very active} \\
  \noalign{\smallskip}

J03473$-$019&G 80-021&YD&AB Dor&AAA.A           &       Y\\     
J04472+206*&RX J0447.2+2038&YD&IC 2391/Argus&AAAAA              &Y\\
J05019+011&1RXS J050156.7+010845&YD&IC 2391/$\beta$ Pic &AAA.A          &Y\\
J05062+046*&RX J0506.2+0439&YD&IC 2391/Argus/$\beta$ Pic&AAA.A          &Y\\
J06000+027  & G 99$-$049  &   YD  &   ... &   AAA.A   &   Y   \\
J06318+414*&LP 205-044&YD&...&AAA.A                                             &Y\\
J06574+740*  & 2MASS J06572616+7405265  &   D   &   ... &   AAAAN   &   F   \\
J07319+362N &  V* BL Lyn     &   YD  &   Castor  &  ANAAA    &   Y   \\ 
J07446+035  & YZ CMi  &   YD  &   IC 2391/Argus   &   AAA.A   &   Y   \\
J07472+503&2MASS J07471385+5020386&D&UMa&AANAA                                          &Y\\
J07558+833&GJ 1101&YD&Hyades SC&AAAAA                                           &Y\\
J09161+018&RX J0916.1+0153&D&UMa&AAA.A                                          &Y\\
J09449$-$123&G 161-071&YD&IC 2391/Argus&AAAAA                                           &Y\\
J10360+051&RY Sex&D&...&ANAAA                                           &F\\
J11201$-$104&LP 733-099&YD&...&AAAAA                                            &Y\\
J11476+002&LP 613-049 A&YD&IC 2391&ANAAA                                                &Y\\
J12156+526*&StKM 2-809&D&UMa&AAAAA                                              &Y\\
J12428+418&G 123-55&YD&Hyades&ANAAA                                             &Y\\
J13005+056&FN Vir&D&...&AAAAA                                           &F\\
J13536+776&RX J1353.6+7737&D&UMa&AAAAA                                          &Y\\
J13591-198  &  LP 799$-$007 &   YD  &   IC 2391 &   AAAAA   &   Y   \\
J14173+454*&RX J1417.3+4525&D&...&AAA.A                                         &F\\
J15218+209&OT Ser&D&...&AAAAA                                           &F\\
J15499+796*&LP 23-35&YD&...&AAANA & Y \\
J16570$-$043&LP 686-027&YD&...&AAAAA                                            &Y\\
J18022+642&LP 071-082&D&UMa&AAAAA                                               &Y\\
J18131+260&LP 390-16&D&...&AAAAA                                                &F\\
J18174+483  &  TYC 3529-1437-1 &   YD  &   ... &   ANAAA   &   Y   \\
J18498-238  & V1216 Sgr  &   D   &   Castor  &   AA.AA   &  Y    \\
J19422-207  &  2MASS J19421282-2045477 &   YD  &   ... &   AAA.A   &   Y   \\
J19511+464&G 208-042&YD&IC 2391/Argus &AAA.A                                            &Y\\
J20451$-$313&AU Mic&YD&$\beta$ Pic &AAAAA                                               &Y\\
J22012+283&V374 Peg&YD&Castor&AAAAA                                             &Y\\
J22468+443  & EV Lac  &   D   &   UMa &   AAA.A   &   Y   \\
J22518+317&GT Peg&YD&...&AAA.A                                          &Y\\
J23548+385&RX J2354.8+3831&D&UMa&ANAAA                                          &Y\\

  \noalign{\smallskip}
  \hline
  \noalign{\smallskip}
  
\multicolumn{6}{c}{\includegraphics[width=2mm]{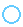} \em Less active} \\
  \noalign{\smallskip}

 J04153$-$076 & $o^{02}$ Eri C & TD-D & ... & ANA.A&F\\
 J05366+112 & 2MASS J05363846+1117487 & YD & ... & AN..A&Y\\
 J07033+346 & LP 255$-$011 & D & ... & AN.AA&F\\
 J10196+198 & AD Leo & YD & ... & ANA.A&Y\\
 J10584$-$107 & LP 731$-$076 & YD & ... & NNNAA&Y\\
 J11055+435 & WX UMa & TD-D & ... & NNA.N&F\\
 J11474+667* & 1RXS J114728.8+664405 & YD & Castor & NNNAA&Y\\
 J12189+111 & GL Vir & YD & Hyades & AAANN&Y\\
 J13102+477 & G 177$-$025 & D & ... & NNAAA&F\\
 J14321+081 & LP 560-035 & YD & Hyades& NN.NN&D\\
 J15305+094 & NLTT 40406 & D & ... & AA.NN&F\\
 J16313+408* & G 180$-$060 & YD & IC 2391/Argus & AN.NN&D\\
 J17338+169* & 1RXS J173353.5+165515 & YD & ... & AAA.N&Y\\
 J18075$-$159 & GJ 1224 & D & ... & ANA.A&F\\
 J18189+661 & LP 71$-$165 & D & ... & AAN.N&F\\
 J20093$-$012 & 2MASS J20091824-0113377 & D & ... & ANA.A&F\\
 J22231-176 &  LP 820$-$012  &   D   &   ... &   NNAAN   &   F   \\
\hline
\end{tabular}
\tablefoot{
\tablefoottext{a}{Asterisk ``*'' in the Karmn identifier marks stars with a new close binary candidate companion \citep{Cifuentes24}}
\tablefoottext{b}{In the five-character activity flag, `A' and `N' indicate whether the star satisfies each of the Eqs.~\ref{eq.prot}--\ref{eq.pew}. Non-literature values for assessment are represented with `.'.
The five labels refer to P$_{\rm{rot}}$, $v\sin{i}$, $\rm{L}_X/\rm{L}_J$, $NUV-J$, and pEW(H$\alpha$) in this order.}
\tablefoottext{c}{Classes `D', `F', and `Y' indicate dubious, field, and young stars, respectively, as established in this work.}
}
\end{table}

\cite{Reiners2022} measured the surface-averaged magnetic fields of 36 of the 118 very active stars.
The 36 M0.0--5.0\,V stars are listed in the top part of Table~\ref{table.27} and are illustrated in Fig.~\ref{fig:magneticfields}.
As expected, all of the 36 stars display strong magnetic fields, of over 1900\,G.
Except in six cases (2MASS J06572616+7405265 / J06574+740, RY~Sex / J10360+051, FN~Vir / J13005+056, RX~J1417.3+4525 / J14173+454, OT~Ser / J15218+209, and LP~390-16 / J1831+260), they belong to the young disc, to an SKG, or both.
Of the six exceptional stars, one (RX~J1417.3+4525) displays a large scatter of precise radial velocities in the CARMENES survey \citep{Ribas23}. It is also a new close binary candidate, previously reported by \cite{Cifuentes24}. It may be another pair of stars with enhanced magnetic activity due to a fast, tidally locked rotation of only 0.37~$\pm$~0.04\,d \citep{Shan24}.
However, some of the other four stars in the thin disc have been repeatedly included as targets for high-contrast surveys for young sub-stellar companions \citep[e.g.][]{Dae07,AllenReid08,Beichman10,Salama21} and are very well characterised.
These 6 stars without young kinematics over the 36 stars with strong magnetic fields represent a fraction of 17\%. Although the sample is not statistically significant, it can be interpreted as a contamination rate that highlights the fundamental role of combining activity with kinematics when assessing youth.

As a result, we may conclude that some young stars can either exhibit kinematics that are inconsistent with the young disc and they are actually older than stars in the young disc, but have retained enhanced activity for longer (or both). 
The opposite case may also be true, namely, that some old stars can either have kinematics consistent with the young disc and, although they are actually younger than stars in the thin disc, they retained enhanced activity for a shorter amount of time (e.g. BD+01 2447 / J10289+008 in the AB~Dor moving group; \citealt{Malo13,Zuck11,Gagne18Banyan} -- but see \citealt{Ama21}). Alternatively, both may again be valid at once.
These conclusions have a tremendous impact on current attempts to derive stellar ages in general, and of M dwarfs that have burned their lithium in particular, as it may be extremely difficult to disentangle stellar activity, kinematic membership, and actual age. Furthermore, this would make it difficult  to derive reliable activity-age relations.

There are another 17 M dwarfs in common between our work and that of \cite{Reiners2022} with magnetic fields above 1900\,G (listed at the bottom of Table~\ref{table.27}):\ here, there are 8 stars of the young disc or a SKG, 7 thin-disc stars, and 2 thick-thin transition disc stars (i.e. $o^{02}$~Eri~C / J04153-076 and WX~UMa / J11055+435). 
We were not able to corroborate the young age for only  1 out of the 8 kinematically young stars (i.e. the fast-rotating M6.0\,V star LP~560-35 / J14321+081) based on Sect.~\ref{sec.activeandyoung}.
These results support the aforementioned stellar activity-kinematic membership-age entangling in late-type stars.

Despite these difficulties, our sample includes a very few obvious cases of significant activity-age relations.
For example, among the 22 M dwarfs classified as young with at least four activity indicators satisfying Eqs.~\ref{eq.prot}--\ref{eq.pew} and with strong magnetic fields (of over 1900\,G), there are 9 stars that are perhaps the youngest, most active, nearest M dwarfs observable from Calar Alto, with ages of less than about 50\,Ma.
This list of very young, very active stars includes AU~Mic / J20451$-$313, 1RXS~J050156.7+010845 / J05019+011 and RX~J0506.2+0439 / J05062+046 in the $\beta$~Pictoris Moving Group \citep{Barrado99,Zuckerman01,AF15betapic}, G~161-071 / J09449$-$123, LP~613-49\,A / J11476+002, RX~J0447.2+2038 / J04472+206, and G~208-042 / J19511+464 in Argus or IC~2391 (\citealt{Malo13}; this work).


\section{Summary}\label{sec.conclusions}

We provide the spectral types, astrometric, and kinematic parameters for 2215 M dwarfs and 3 late K dwarfs contained in the Carmencita database. We report the radial velocities for 2187 of them, which we used to compute their $UVW$ galactocentric velocities.

On the one hand, we placed these stars in the different galactic populations: galactic halo (1), thick disc (150), transition between the thick and thin disc (78), and thin disc (1958). Among the thin-disc population, we distinguished sources belonging to the young disc based on the definition of \cite{Leg92} (713).
On the other hand, we used {\tt SteParKin}, {\tt LACEwING}, and {\tt BANYAN} $\Sigma$ algorithms to assign the stars to any young SKG in the solar neighbourhood, based on kinematics.

To assess their youth, we evaluated the dependence of stellar rotation and activity with $G-J$ colour as a proxy of spectral type and age through X-rays, H$\alpha$, and $NUV$ {\em GALEX} band emission, along with the rotational velocity and rotational period by using members of the Praesepe and Hyades clusters, as well as bona fide members of young SKGs in our sample. We observe that young (<800\,Ma) M dwarfs are typically constrained within certain rotation and activity limits, providing evidence for fast rotation and strong chromospheric and coronal emissions in young low-mass stars, in comparison with older sources. We  provide five rotation- and activity-colour relations in the M dwarf regime that young stars are expected to satisfy.
We applied these equations to 1546 M dwarfs without known companions and with a colour range of 1.9\,mag < $G-J <$ 3.8\,mag. We defined two indices to help us quantify the activity and kinematic results. We identified 191 young M dwarfs, 113 of which are  presented for the first time as young in the framework of this paper. Another 87 stars have been identified as potentially young, whereas for 99 stars, the classification is not fully clear (dubious) and for 173 stars, we do not have  enough data for carry out a proper assessment (inconclusive). Finally, 996 stars have been identified as field members. Furthermore, we also identified the 118 most active sources in our sample regardless their kinematics. Of them, 97 are young stars as by our criteria.
Out of the 118 very active M dwarfs, there are 36 stars with strong magnetic fields over 1900\,G, among which we identified nine M dwarfs likely younger than 50\,Ma. We identified another 17 stars from the sample of 1546 stars without outstanding levels of activity and with magnetic fields over 1900\,G.

Previous searches for M dwarfs that are either very young (very active or in SKGs) or very old (in the thick disc or halo) have been quite complete, as very few young nearby stars in SKGs still lack  assignations.
However, there is much more work that needs to be done in terms of determining more precise ages for M dwarfs.
For that purpose, astronomers will have to continue battling on several fronts.
One is to go on collecting, improving, and simultaneously analysing  all kinematic, photometric (rotation periods, and UV flux excess), spectroscopic (rotational velocity, H$\alpha$, and other chromospheric lines), and X-ray data.
The other approach is to go on measuring such activity indicators of late-type stars in open clusters and SKGs with well-determined ages. Thus, the more clusters, stars per cluster, and different cluster ages at our disposal, the better the results we will be able to obtain in the future. 

\section{Data availability}

Long tables~\ref{table.name_coords_pm_d}, \ref{table.name_vr_uvw}, \ref{table.name_activity_params}, and \ref{table.name_pop_skg} are only available in electronic form at the CDS via anonymous ftp to {\tt cdsarc.u-strasbg.fr} (130.79.128.5) or via {\tt http://cdsweb.u-strasbg.fr./cgi-bin/qcat?J/A+A/*/*}}.

\begin{acknowledgements}

We thank the anonymous referee for their constructive comments that helped to improve the quality of this paper.

We also acknowledge A.\,J.~Dom\'inguez-Fern\'andez, I.~Gallardo, and L.~Peralta de~Arriba for their valuable assessment, and R.~Clavero and D.~Jones for the ALFOSC observations.

 CARMENES is an instrument at the Centro Astron\'omico Hispano en Andaluc\'ia (CAHA) at Calar Alto (Almer\'{\i}a, Spain), operated jointly by the Junta de Andaluc\'ia and the Instituto de Astrof\'isica de Andaluc\'ia (CSIC).
   CARMENES was funded by the Max-Planck-Gesellschaft (MPG), the Consejo Superior de Investigaciones Cient\'{\i}ficas (CSIC),  the Ministerio de Econom\'ia y Competitividad (MINECO) and the European Regional Development Fund (ERDF) through projects FICTS-2011-02, ICTS-2017-07-CAHA-4, and CAHA16-CE-3978, and the members of the CARMENES Consortium 
 with additional contributions by the MINECO, 
 the Deutsche Forschungsgemeinschaft (DFG) through the Major Research Instrumentation Programme and Research Unit FOR2544 ``Blue Planets around Red Stars'', 
 the Klaus Tschira Stiftung, the states of Baden-W\"urttemberg and Niedersachsen, and by the Junta de Andaluc\'{\i}a.

 This article is based on observations made in the Observatorios de Canarias del IAC with the Nordic Optical Telescope (NOT) operated on the island of La Palma in the Observatorio del Roque de los Muchachos through programme number 64-264. 
 
 We acknowledge financial support from the Agencia Estatal de Investigaci\'on (AEI/10.13039/501100011033) of the Ministerio de Ciencia, Innovaci\'on y Universidades and the ERDF ``A way of making Europe'' through projects 
 PID2022-137241NB-C4[1:4],      
 PID2021-125627OB-C31,          
 and the Centre of Excellence ``Severo Ochoa'' and ``Mar\'ia de Maeztu'' awards to the Instituto de Astrof\'isica de Canarias (CEX2019-000920-S), Instituto de Astrof\'isica de Andaluc\'ia (CEX2021-001131-S) and Institut de Ci\`encies de l'Espai (CEX2020-001058-M).

We also acknowledge support from the ``Tecnolog\'ias avanzadas para la exploraci\'on de universo y sus componentes'' (PR47/21 TAU) project funded by Comunidad de Madrid, by the Recovery, Transformation and Resilience Plan from the Spanish State, and by NextGenerationEU from the European Union through the Recovery and Resilience Facility.

This research made use of the CDS cross-match service \citep{CDSBoch12,CDSPineau2020}, 
SIMBAD database \citep{Wenger00}, 
VizieR catalogue access tool \citep{Och00}, 
and Aladin sky atlas \citep{Bonn00, BF14} provided by CDS, Strasbourg, France. 
This research also made use of the TOPCAT \citep{Taylor05}, STILTS \citep{Taylor06}, and the SVO Discovery Tool, developed by the Spanish Virtual Observatory \url{https://svo.cab.inta-csic.es} project funded by MCIN/AEI/10.13039/501100011033/ through grant PID2020-112949GB-I00.

\end{acknowledgements}

\bibliographystyle{aa} 
\bibliography{biblio} 



\begin{appendix}


\onecolumn

\section{Additional tables}

\label{sec.data}

\begin{table*}
  \centering
  \caption {References for distances, proper motions, and radial velocities.}
  \label{table.3_ref}
   \small
  \begin{tabular}{l l c}
  \hline 
  \hline
  \noalign{\smallskip}
Catalogue       & Reference     &       No. of stars  \\

  \noalign{\smallskip}
  \hline
  \noalign{\smallskip}
  
\multicolumn{3}{c}{\it Distances} \\
  \noalign{\smallskip}

{\em Gaia} DR3  &       \cite{GaiaDR3}  &        2079   \\
          \noalign{\smallskip}
{\em Gaia} DR2  &       \cite{GaiaDR2}  &       51              \\
          \noalign{\smallskip}
This work (spectro-photometric) &       ...     &       24      \\
      \noalign{\smallskip}
{\em Hipparcos}, the new reduction      &       \cite{HIP2}     &       18              \\
      \noalign{\smallskip}
Trigonometric parallaxes for 1507 nearby mid-to-late M dwarfs   &       \cite{Ditt14}   &       16                 \\
      \noalign{\smallskip}
Yale trigonometric parallaxes, fourth edition   &       \cite{vAl95cat} &       10                         \\
          \noalign{\smallskip}
Others (parallactic)    &       Various\tablefootmark{a}        &        7       \\
          \noalign{\smallskip}    
1103 parallaxes and proper motions from URAT & \cite{FZ16} & 8 \\
          \noalign{\smallskip}
CARMENES input catalogue of M dwarfs. II. High-resolution imaging... & \cite{Cor17} & 3 \\

          \noalign{\smallskip}
The Gould's Belt distances survey (GOBELINS). IV. Distance, depth...      & \cite{Galli18}        &       2       \\        
  \noalign{\smallskip}
  \hline
  \noalign{\smallskip}

\multicolumn{3}{c}{\it Proper motions} \\
  \noalign{\smallskip}
  
{\em Gaia} DR3  &       \cite{GaiaDR3}  &        2079           \\
          \noalign{\smallskip}
{\em Gaia} DR2 & \cite{GaiaDR2} & 51 \\
      \noalign{\smallskip}
The PPMXL catalog       &       \cite{Roeser10} &       51      \\ 
      \noalign{\smallskip}
This work (astrometric fit)     &       ...     &       22      \\ 
      \noalign{\smallskip}
{\em Hipparcos}, the new reduction      & \cite{HIP2}   &       14      \\
          \noalign{\smallskip}    
WISE J072003.20-084651.2B is a Massive T Dwarf& \cite{Dupuy19}  &       1       \\

  \noalign{\smallskip}
  \hline
  \noalign{\smallskip}

\multicolumn{3}{c}{\it Radial velocities} \\
  \noalign{\smallskip}
{\em Gaia} DR3  &       \cite{GaiaDR3}  &       1861            \\
          \noalign{\smallskip}
Palomar/MSU nearby star spectroscopic survey & \cite{Hawley96}  & 80 \\
  \noalign{\smallskip}

Chromospheric Activity of M Stars Based on LAMOST Low- and Medium... & \cite{Zhang21}  &  46 \\

\noalign{\smallskip}
  {\em Gaia} EDR3 &  \cite{GaiaeDR3} &  34 \\
         \noalign{\smallskip}

CARMENES input catalogue of M dwarfs. III. Rotation and activity from... & \cite{Jeff18}&28\\

   \noalign{\smallskip}
The CARMENES search for exoplanets around M dwarfs. Radial velocities... & \cite{Lafarga20} & 25 \\

    \noalign{\smallskip}
 \noalign{\smallskip}
   Near-infrared metallicities, radial velocities, and spectral types for 447...     &       \cite{New14}    &       24\\
   
Others  &       Various\tablefootmark{b}        &       22 \\
 \noalign{\smallskip}

Thirty New Low-mass Spectroscopic Binaries& \cite{Shk10}  &  10 \\
          \noalign{\smallskip}
  Stellar Atmospheric Parameters of M-type Stars from LAMOST DR8 & \cite{Ding22}&10\\
          \noalign{\smallskip}

The CARMENES search for exoplanets around M dwarfs. Spectroscopic... & \cite{Baro21} & 9 \\
  \noalign{\smallskip}
   A Near-Infrared Spectroscopic Survey of 886 Nearby M Dwarfs  &  \cite{Terrien15}  &  8  \\
\noalign{\smallskip}
The CARMENES search for exoplanets around M dwarfs. Nine new... & \cite{Baro18} &  8\\

\noalign{\smallskip}

{\em Gaia} DR2  &       \cite{GaiaDR2}  &       8               \\
        
          \noalign{\smallskip}

Intermediate Resolution Near-infrared Spectroscopy of 36 Late M Dwarfs & \cite{Des12} & 7 \\

          \noalign{\smallskip}

ACRONYM. III. Radial Velocities for 336 Candidate Young Low-mass... & \cite{Schn19}&6\\
 \noalign{\smallskip}

This work (CARMENES) &  ... &  1  \\
 
  \noalign{\smallskip}
  \hline
  \end{tabular}
\tablefoot{
  \tablefoottext{a}{\cite{Dupuy19,Fin18,Hen06,Lep09,Ried10,Wein16}.}
  \tablefoottext{b}{Other references on $V._r$ -- \cite{Burg15,Gizis2002,Hal18,Kou19,Malo14,Marfil21,New16,Per17,kunder17,RB09,Shk12,Sou18,Sper16,Wilson67}.}
}
\end{table*}

\begin{table*}[!ht]
  \centering
  \caption {References for rotational periods, rotational velocities, and H$\alpha$ pseudoequivalent widths$^d$.}
  \label{tab.indicatorsrefs}
  \small
  \begin{tabular}{l l c}
  \hline 
  \hline
  \noalign{\smallskip}
Catalogue       & Reference     &       No. of stars  \\

  \noalign{\smallskip}
  \hline
  \noalign{\smallskip}
  
\multicolumn{3}{c}{\it Rotational periods} \\
  \noalign{\smallskip}

CARMENES input catalogue of M dwarfs. IV. New rotation periods... &  \cite{DA19}  &  98 \\
          \noalign{\smallskip}

This work (TESS, K2) & ...  &  93 \\
          \noalign{\smallskip}

CARMENES input catalog of M dwarfs. VII. New rotation periods...&\cite{Shan24} &  72 \\
          \noalign{\smallskip}

Others & Various$^b$ & 65  \\
          \noalign{\smallskip}

A Photometric Variability Survey of Field K and M Dwarf Stars...  &  \cite{Har11}  &  44  \\
          \noalign{\smallskip}

The Rotation and Galactic Kinematics of Mid M Dwarfs... &  \cite{New16}  &  31 \\
          \noalign{\smallskip}

An Activity-Rotation Relationship and Kinematic Analysis...  &  \cite{West15}  &  30  \\
          \noalign{\smallskip}

New Rotation Period Measurements for M Dwarfs in the Southern...  &  \cite{New18}  &  24  \\
          \noalign{\smallskip}

HADES RV programme with HARPS-N at TNG. VII Rotation and activity...  &  \cite{SM18} &  20  \\
          \noalign{\smallskip}

On the Angular Momentum Evolution of Fully Convective Stars... &  \cite{Irw11}  &  13     \\
          \noalign{\smallskip}

Rotation-activity relations and flares of M dwarfs with K2... &  \cite{Rae20}  &  12  \\
          \noalign{\smallskip}

The CARMENES search for exoplanets around M dwarfs. Period search... & \cite{Fuhr19}  &  11   \\

  \noalign{\smallskip}
  \hline
  \noalign{\smallskip}
  
\multicolumn{3}{c}{\it Rotational velocities} \\
  \noalign{\smallskip}

The CARMENES search for exoplanets around M dwarfs. High-resolution... &  \cite{Rein18}  &  177  \\

          \noalign{\smallskip}

Magnetism, rotation, and nonthermal emission in cool stars. Average magnetic...& \cite{Reiners2022} &   133   \\

          \noalign{\smallskip}

CARMENES input catalogue of M dwarfs. III. Rotation and activity... & \cite{Jeff18}  &  110  \\

          \noalign{\smallskip}

Others & Various$^c$   &  45   \\
          \noalign{\smallskip}

The Rotation of M Dwarfs Observed by the Apache Point Galactic Evolution... & \cite{Gilh18}  &  37  \\
          \noalign{\smallskip}

A Catalog of Rotation and Activity in Early-M Stars & \cite{Rein12}  &  30  \\
          \noalign{\smallskip}

Magnetic Inflation and Stellar Mass. II. On the Radii of Single, Rapidly Rotating... & \cite{Kess18}  &  25  \\

          \noalign{\smallskip}

Rotational Velocities for M Dwarfs& \cite{Jen09}  &  23  \\  
          \noalign{\smallskip}

Observation and modelling of main-sequence star chromospheres -- XIV. Rotation...& \cite{Hou10}  &  19   \\
          \noalign{\smallskip}

SPIRou Input Catalogue: global properties of 440 M dwarfs observed with... &  \cite{Fou18}  &  16   \\
          \noalign{\smallskip}

A Spectroscopic Survey of a Sample of Active M Dwarfs &  \cite{Moc02}  &  11   \\

  \noalign{\smallskip}
  \hline
  \noalign{\smallskip}
  
\multicolumn{3}{c}{\it H$\alpha$ pseudoequivalent widths} \\
  \noalign{\smallskip}

Trumpeting M dwarfs with CONCH-SHELL: a catalogue of nearby cool...
& \cite{Gai14}  &  614 \\
          \noalign{\smallskip}

The CARMENES search for exoplanets around M dwarfs. Diagnostic... & \cite{Fuhr22}  &  292  \\
          \noalign{\smallskip}

CARMENES input catalogue of M dwarfs. I. Low-resolution spectroscopy... & \cite{AF15} &  173 \\
          \noalign{\smallskip}

Magnetic Activities of M-type Stars Based on LAMOST DR5 and Kepler...& \cite{Lu19} &  86  \\
          \noalign{\smallskip}

The H$\alpha$ Emission of Nearby M Dwarfs and its Relation to Stellar Rotation & \cite{New17}  &  81  \\
          \noalign{\smallskip}

Others & Various$^d$  &  48  \\
          \noalign{\smallskip}

The Palomar/MSU Nearby-Star Spectroscopic Survey. I. The Northern... & \cite{Reid95}  &  32 \\
          \noalign{\smallskip}

Activity of M dwarfs in the CARMENES sample & \cite{Schf15}  &  30  \\
          \noalign{\smallskip}

The CARMENES search for exoplanets around M dwarfs. Activity indicators... & \cite{Schf19}  &  23  \\
          \noalign{\smallskip}

The Palomar/MSU Nearby Star Spectroscopic Survey. III. Chromospheric... & \cite{Gizis2002}  &  10 \\

  \noalign{\smallskip}
  \hline
  \end{tabular}
\tablefoot{
  \tablefoottext{a}{References for $NUV-J$ are \citet{Skru06} and \citet{GALEX}, and for $L_X / L_J$ is \citet{Rosat} and this work.}
  \tablefoottext{b}{Other references on $P$ -- \cite{Koe02,Poj02,Wat06,Nor07,Mori08,Berh10,Har10,Mor10,Kira12,KS13,McQ13,Bidd14,SM15,Stelzer16,SM17b,Luque18,Diaz19,TP19,Bluhm20,Dreizler20,Med20,Stef20,Sto20_periods,Ama21,Ded21,Lafa21,Skr21,Luque22,Fou23,Irv23,SM23}.}
  \tablefoottext{c}{Other references on $v \sin{i}$ -- \cite{MC92,Jefr95,Tok92,Sta97,Gizis2002,Reid02,MB03,Torr06,Whi07,Bro10,LS10,MA10,RB10,Schl10,Schlieder12b,Des12,Barn14,Malo14,MR14,Stef20,Marfil21}.}
  \tablefoottext{d}{Other references on pEW(H$\alpha$) -- \cite{Boch05,Silvestri05,PB06,Riaz06,Schm07,Shk09,RB10,Lep13,Kraus14,MR14,Ried14,West15,Jon16,Schf19,Med20,Fuhr23}.}
}
\end{table*}

\section{Long tables}

\input{TA1_2218_v1mod}
\input{TA2_2187}
\input{TA3_1581_v1mod}
\input{TA4_2187_v0mod}

\end{appendix}

\end{document}

%% file: TA2_2187.tex
\centering
\begin{longtable}{lcclcccccc}
\label{table.name_vr_uvw}\\
\caption{Carmencita identifier, radial velocities, and Galactic space velocities computed in this work.}\\
	\noalign{\smallskip}
\hline \hline
	\noalign{\smallskip}
Karmn & $V_r$ &   $\delta V_r$    &   Ref.\tablefootmark{a}   &   $U$   &   $\delta U$  &   $V$   &   $\delta V$  &  $W$   &   $\delta W$ \\
    &   [km\,s$^{-1}$] &   [km\,s$^{-1}$] &   &   [km\,s$^{-1}$] &   [km\,s$^{-1}$] &   [km\,s$^{-1}$] &   [km\,s$^{-1}$] &   [km\,s$^{-1}$] &   [km\,s$^{-1}$] \\
	\noalign{\smallskip}

	\hline
	\noalign{\smallskip}
	\endfirsthead
\caption{Carmencita identifier, radial velocities, and Galactic space velocities computed in this work (continued).}\\
\hline \hline
	\noalign{\smallskip}
Karmn & $V_r$ &   $\delta V_r$    &   Ref.\tablefootmark{a}   &   $U$   &   $\delta U$  &   $V$   &   $\delta V$  &  $W$   &   $\delta W$ \\
    &   [km\,s$^{-1}$] &   [km\,s$^{-1}$] &   &   [km\,s$^{-1}$] &   [km\,s$^{-1}$] &   [km\,s$^{-1}$] &   [km\,s$^{-1}$] &   [km\,s$^{-1}$] &   [km\,s$^{-1}$] \\
 		\noalign{\smallskip}
	\hline
	\noalign{\smallskip}
	\endhead
	\hline
	\noalign{\smallskip}
\endfoot
  J00012+139N & -6.073 & 0.383 & DR3 & -13.143 & 0.135 & 10.564 & 0.292 & 20.898 & 0.322\\
  J00012+139S & -10.255 & 0.163 & DR3 & -13.83 & 0.03 & 6.746 & 0.108 & 23.338 & 0.12\\
  J00026+383 & 1.358 & 1.262 & DR3 & 13.004 & 0.456 & 6.012 & 1.071 & -2.089 & 0.506\\
  J00033+046 & -6.005 & 0.276 & DR3 & 9.137 & 0.03 & -9.775 & 0.151 & -0.384 & 0.229\\
  J00051+457 & -0.792 & 0.139 & DR3 & -38.744 & 0.056 & -23.778 & 0.121 & -16.029 & 0.039\\
  J00056+458 & -2.239 & 0.138 & DR3 & -37.079 & 0.057 & -24.481 & 0.12 & -15.19 & 0.039\\
  J00067-075 & -40.458 & 0.298 & DR3 & 38.222 & 0.008 & -41.713 & 0.113 & 25.985 & 0.276\\
  J00077+603 & 1.529 & 0.974 & DR3 & -21.097 & 0.452 & -9.451 & 0.863 & -4.236 & 0.036\\
\hline
\end{longtable}
\tablefoot{
    \tablefoottext{a}{Bar18: \cite{Baro18}; Bar21: \cite{Baro21}; Burg15: \cite{Burg15}; D22: \cite{Ding22}; DR2: \cite{GaiaDR2}; DR3: \cite{GaiaDR3}; Des12: \cite{Des12}; Dy54: \cite{Dy54}; EDR3: \cite{GaiaeDR3}; Giz02: \cite{Gizis2002}; Hal18: \cite{Hal18}; Haw96: \cite{Reid95}; Jeff18: \cite{Jeff18}; Kou19: \cite{Kou19}; Lafa20: \cite{Lafarga20}; Malo14: \cite{Malo14}; Marf21: \cite{Marfil21}; New14: \cite{New14}; New16: \cite{New16}; Per17: \cite{Per17}; RAVE5: \cite{kunder17}; RB09: \cite{RB09}; Reid95: \cite{Reid95}; Schn19: \cite{Schn19}; Shk10: \cite{Shk10}; Shk12: \cite{Shk12}; Sou18: \cite{Sou18}; Ter15: \cite{Terrien15}; Wil67: \cite{Wilson67}; Z21: \cite{Zhang21RV}.}

}

%% file: TA3_1581_v1mod.tex
\centering
\begin{landscape}
\begin{longtable}{llclclclcc}
\label{table.name_activity_params}\\
\caption{Carmencita identifier and compiled activity parameters.}\\
	\noalign{\smallskip}

\hline \hline
	\noalign{\smallskip}
Karmn	 & 	   $v\sin{i}$	 & 	 $v\sin{i}$ 	&	Ref.\tablefootmark{a}	&  $P$	 	 & 	 Ref.\tablefootmark{b}	& pEW(H$\alpha$)	 & 	Ref.\tablefootmark{c}	&	$\log{L_{\rm X}/L_J}$\tablefootmark{d}	&       $NUV-J$\tablefootmark{e}	           \\
	&		flag		&	[kms$^{-1}$] 	&				&	[d]	&				&	[\AA]				&				&		&						[mag]					\\
	\noalign{\smallskip}

	\hline
	\noalign{\smallskip}
	\endfirsthead

\caption{Carmencita identifier and compiled activity parameters (continued).}\\
\hline \hline
	\noalign{\smallskip}
Karmn	 & 	   $v\sin{i}$	 & 	 $v\sin{i}$ 	&	Ref.\tablefootmark{a}	&  $P$	 	 & 	 Ref.\tablefootmark{b}	& pEW(H$\alpha$)	 & 	Ref.\tablefootmark{c}	&	$\log{L_{\rm X}/L_J}$\tablefootmark{d}	&       $NUV-J$\tablefootmark{e}	           \\
	&		flag		&	[kms$^{-1}$] 	&				&	[d]	&				&	[\AA]				&				&		&						[mag]					\\

		\noalign{\smallskip}
	\hline
	\noalign{\smallskip}
	\endhead
	\hline
	\noalign{\smallskip}
\endfoot
J00012+139S&...&...&...&...&...&-0.41&Lep13&...&11.498\\
J00033+046&...&...&...&...&...&0.41$\pm$0.08&Gai14&...&12.348\\
J00051+457&<&2.0&Rein18&15.37$\pm$0.09&DA19&0.347$\pm$0.018&Fuhr22&...&...\\
J00067-075&<&2.0&Rein18&126.0$\pm$15.0&SM23&-0.057$\pm$0.069&Fuhr22&...&...\\
J00079+080&...&...&...&...&...&-1.782$\pm$0.424&Lu19&...&...\\
J00084+174&...&...&...&...&...&0.4$\pm$0.07&Gai14&...&11.433\\
J00110+052&...&...&...&...&...&0.48$\pm$0.06&Gai14&...&12.055\\
J00115+591&...&...&...&...&...&-1.6$\pm$0.4&AF15a&...&...\\
\hline
\end{longtable}
\tablefoot{
\tablefoottext{a}{Barn14: \cite{Barn14}; Bro10 \cite{Bro10}; Des12: \cite{Des12}; Fou18: \cite{Fou18}; Gilh18: \cite{Gilh18}; Giz02: \cite{Gizis2002}; Hou10: \cite{Hou10}; Jeff18: \cite{Jeff18}; Jefr95: \cite{Jefr95}; Jen09: \cite{Jen09}; Kess18: \cite{Kess18}; LS10: \cite{LS10}; MA10: \cite{MA10}; MB03: \cite{MB03}; MC92: \cite{MC92}; MR14: \cite{MR14}; Malo14: \cite{Malo14}; Marf21: \cite{Marfil21}; Moc02: \cite{Moc02}; RB10: \cite{RB10}; Reid02: \cite{Reid02}; Rein12: \cite{Rein12}; Rein18: \cite{Rein18}; Rein22: \cite{Reiners2022}; Schl10: \cite{Schl10}; Schl12b: \cite{Schlieder12b}; Sta97: \cite{Sta97}; Stef20: \cite{Stef20}; Tok92: \cite{Tok92}; Tor06: \cite{Torr06}; Whi07: \cite{Whi07}.}
\tablefoottext{b}{Ama21: \cite{Ama21}; Berh10: \cite{Berh10}; Bidd14: \cite{Bidd14}; Blu20: \cite{Bluhm20}; DA19: \cite{DA19}; Ded21: \cite{Ded21}; Dre20: \cite{Dreizler20}; Fou23: \cite{Fou23}; Hart10: \cite{Har10}; Hart11: \cite{Har11}; Irv23: \cite{Irv23}; Irw11: \cite{Irw11}; KE02: \cite{Koe02}; KS13: \cite{KS13}; Kira12: \cite{Kira12}; Lafa21: \cite{Lafa21}; Luq18: \cite{Luque18}; Luq22: \cite{Luque22}; McQ13: \cite{McQ13}; Med20: \cite{Med20}: Mori08: \cite{Mori08}; Mori10: \cite{Mor10}; New16: \cite{New16}; New18: \cite{New18}; Nor07: \cite{Nor07}; Poj02: \cite{Poj02}; Rae20: \cite{Rae20}; Rev20: \cite{Rev20}; SM15: \cite{SM15}; SM17b: \cite{SM17b}; SM18: \cite{SM18}; SM23: \cite{SM23}; Shan24: \cite{Shan24}; Skr21: \cite{Skr21}; Ste16: \cite{Stelzer16}; Stef20: \cite{Stef20}; Sto20: \cite{Sto20_periods}; TP19: \cite{TP19}; Wat06: \cite{Wat06}; West15: \cite{West15}.}
\tablefoottext{c}{AF15a: \cite{AF15}; Boch05: \cite{Boch05}; Fuhr22: \cite{Fuhr22}; \cite{Fuhr23}; Gai14: \cite{Gai14}; Giz02: \cite{Gizis2002}; Haw97: \cite{Reid95}; Jon16: \cite{Jon16}; Kra14: \cite{Kraus14}; Lep13: \cite{Lep13}; Lu19: \cite{Lu19}; MR14: \cite{MR14}; Med20: \cite{Med20}; New17: \cite{New17}; PB06: \cite{Phan06}; RB10: \cite{RB10}; Ria06: \cite{Riaz06}; Ried14: \cite{Ried14}; Sche15: \cite{Schf15}; Sche19: \cite{Schf19}; Schn19: \cite{Schn19}; Schm07: \cite{Schm07}; Shk09: \cite{Shk09}; West15: \cite{West15}.}
\tablefoottext{d}{X-rays flux from 1RXS: \cite{Rosat}.}
\tablefoottext{e}{J magnitude from 2MASS \citep{Skru06} and NUV magnitude from GALEX \citep{GALEX}}.
}
\end{landscape}

%% file: TA4_2187_v0mod.tex
\centering
\begin{longtable}{llllllll}
\label{table.name_pop_skg}\\
\caption{Carmencita identifier, stellar population and kinematic groups, and activity flags of the 1647 studies stars. }\\
	\noalign{\smallskip}

\hline \hline
	\noalign{\smallskip}
Karmn	 & 	Pop.\tablefootmark{a}	&	\multicolumn{3}{c}{SKG}	& Activity	&		Close & Final\\
&&&&&flag\tablefootmark{b}&binarity\tablefootmark{c}&assignment\tablefootmark{d}\\
	\noalign{\smallskip}\cline{3-5}	\noalign{\smallskip}
	&		&	{\tt SteParKin}	&	{\tt LACEwING} 	&	{\tt BANYAN} $\Sigma$ 	&&&\\
	&&&(Prob. [\%])&(Prob. [\%])&&\\
	\noalign{\smallskip}
	\hline
	\noalign{\smallskip}
	\endfirsthead
\caption{Carmencita identifier, stellar population and kinematic groups, and activity flags of the 1647 studies stars (continued). }\\
\hline \hline
	\noalign{\smallskip}

Karmn	 & 	Pop.\tablefootmark{a}	&	\multicolumn{3}{c}{SKG}	& Activity	&		Close & Final\\
&&&&&flag\tablefootmark{b}&binarity\tablefootmark{c}&assignment\tablefootmark{d}\\
	\noalign{\smallskip}\cline{3-5}	\noalign{\smallskip}
	&		&	{\tt SteParKin}	&	{\tt LACEwING} 	&	{\tt BANYAN} $\Sigma$ 	&&&\\

	\hline
	\noalign{\smallskip}
	\endhead
	\hline
	\noalign{\smallskip}
\endfoot

    \noalign{\smallskip}

  J00012+139N & D & ... & ... & ... & -- & -- & --\\
  J00012+139S & D & ... & ... & ... & ...AA & ... & F\\
  J00026+383 & D & UMa & ... & ... & -- & -- & --\\
  J00033+046 & YD & ... & ... & ... & ...NN & ... & I\\
  J00051+457 & YD & Hyades & ... & ... & AN..N & ... & D\\
  J00056+458 & YD & Hyades & ... & ... & -- & -- & --\\
  J00067-075 & D & ... & ... & ... & NN..N & ... & F\\
  J00077+603 & YD & IC 2391 & Argus (45) & Argus (94.98) & -- & -- & --\\
\hline
\end{longtable}
\tablefoot{
    \tablefoottext{a}{Galactic kinematic population --
D: Thin disk;
TD: Thick disc;
TD-D: Thick-thin transition disc;
YD: Young disc.}
    \tablefoottext{b}{Activity flag --
``A'' indicates that the source satisfies Eqs.~\ref{eq.xrays}-\ref{eq.prot}, ``N'' otherwise, and ``.'' indicates parameters without any literature value.
The five labels refer to $P_{\rm rot}$, $v\sin{i}$, $L_{\rm X}/L_J$, $NUV-J$, and pEW(H$\alpha$) in this order.}
    \tablefoottext{c}{Close binarity --
y: Reported close binary companion;
y?: Close binary companion candidate.}
    \tablefoottext{d}{Final assignment --
``Y'': Young;
``P'': Potentially;
``D'': Dubious;
``I'': Inconclusive;
``Y'': Young.}
    \tablefoottext{e}{Known young stars candidates --
J00395+149S: Hyades \citep{Roeser19};
J00548+275: 60-300\,Ma \citep{Shk09}, Hyades \citep{Shk12};
J01352-072: 40-300Ma \citep{Shk09}, $\beta$ Pic \citep{AF15betapic}, Tuc-Hor \citep{Kraus14};
J02088+494: 20-300\,Ma \citep{Shk09}, AB Dor \citep{Shk12};
J02186+123: 20-150\,Ma \citep{Shk09};
J02190+238: 35-300\,Ma \citep{Shk09}, Argus \citep{Gagne15};
J02234+227: $\beta$ Pic \citep{AF15betapic};
J02256+375: Hyades \citep{Dou14};
J02412-045: Argus \citep{Bartlett17};
J02519+224: $\beta$ Pic \citep{Gagne18Banyan};
J03147+114: weak-line T~Tauri star candidate \citep{Li2000};
J03186+326: Hyades \citep{Dou14};
J03224+271: Hyades \citep{Roeser19};
J03332+462: AB Dor \citep{Fernandez08};
J03416+552: Columba \citep{Zuck11};
J03463+262: Hyades \citep{Perryman98};
J03473-019: 30-50\,Ma \citep{Shk09}, AB Dor \citep{Zuck04};
J03510+142: $\beta$ Pic \citep{Gagne15};
J03548+163: Hyades \citep{Crain86};
J03565+319: Hyades \citep{Dou14};
J04148+277: Hyades \citep{Johnson62};
J04173+088: AB Dor \citep{Bell15};
J04224+036: Hyades \citep{Dou14};
J04227+205: Hyades \citep{Weis82};
J04238+149: Hyades \citep{Weis79};
J04252+080N: Hyades \citep{vanAltena66};
J04369-162: Tuc-Hor \citep{Kraus14};
J04373+193: Hyades \citep{Griffin88};
J04595+017: $\beta$ Pic \citep{McCarthy12};
J05019+011: $\beta$ Pic \citep{AF15betapic};
J05062+046: $\beta$ Pic \citep{AF15betapic};
J05111+158: Hyades \citep{Delo11};
J05339-023: $\beta$ Pic \citep{AF15betapic};
J05415+534: Hercules-Lyra \citep{Fuhrmann08};
J05599+585: 40-300\,Ma \citep{Shk09}, Castor \citep{Shk12}, AB Dor? \citep{Schlieder12a};
J06075+472: AB Dor \citep{Schlieder12a};
J06310+500: 20-150\,Ma \citep{Shk09};
J06318+414: AB Dor? \citep{Schlieder12a};
J07310+460: Columba \citep{Malo13};
J07319+362N: 25-300\,Ma \citep{Shk09}, Castor \citep{Cab10};
J07384+240: 20-300\,Ma \citep{Shk09}, IC 2391 \citep{Shk12};
J07393+021: Local Association? \citep{Mald10};
J07446+035: $\beta$ Pic \citep{AF15betapic};
J07558+833: Hyades \citep{Eggen93};
J09144+526: Lithium \citep{Bischoff20};
J09248+306: AB Dor? \citep{Schlieder12a};
J09449-123: Argus \citep{Malo13};
J10043+503: 20-300\,Ma \citep{Shk09}, AB Dor? \citep{Schlieder12a};
J10122-037: AB Dor \citep{Malo13};
J10196+198: 25-300\,Ma \citep{Shk09}, Castor \citep{Cab10};
J10289+008: AB Dor \citep{Zuck11};
J10359+288: $\beta$ Pic \citep{Schlieder12a};
J10584-107: TW Hya \citep{Gagne15};
J11031+366: 25-300\,Ma \citep{Shk09};
J11159+553: 15-150\,Ma \citep{Shk09}, Pleiades Supercluster \citep{Shk12};
J11201-104: Argus \citep{Malo13};
J11240+381: 40-300\,Ma \citep{Shk09}, UMa? \citep{Shk12};
J11485+076: 25-300\,Ma \citep{Shk09}, UMa? \citep{Shk12};
J12198+527: AB Dor \citep{Malo13};
J12294+229: 35-300\,Ma \citep{Shk09};
J12417+567: 20-150\,Ma; \citep{Shk09}, UMa? \citep{Shk12};
J13095+289: 60-300\,Ma \citep{Shk09}, Carina \citep{Shk12};
J13294-143: 20-150\,Ma \citep{Shk09};
J13536+776: UMa \citep{capis24};
J13591-198: Argus \citep{Malo13};
J14200+390: young? \citep{Moc02};
J15597+440: AB Dor \citep{Malo13};
J16102-193: USco \citep{Pre01};
J18022+642: 90-300\,Ma \citep{Shk09};
J18451+063: Carina \citep{Elliott14};
J18519+130: 60-300\,Ma \citep{Shk09}, AB Dor? \citep{Schlieder12a};
J20435+240: $\beta$ Pic \citep{Crundall19};
J20451-313: $\beta$ Pic \citep{Cab09};
J21076-130: $\beta$ Pic \citep{Malo13};
J21100-193: $\beta$ Pic \citep{Malo13};
J21185+302: $\beta$ Pic \citep{AF15betapic};
J22012+283: Castor \citep{Cab10};
J22468+443: 25-300\,Ma \citep{Shk09};
J22518+317: 20-300\,Ma \citep{Shk09}, IC 2391? \citep{Lafreniere07};
J23060+639: 30-50\,Ma \citep{Shk09}, AB Dor \citep{Zuck04};
J23083-154: Castor \citep{Cab10};
J23317-027: $\beta$ Pic \citep{AF15betapic};
J23548+385: 25-300\,Ma \citep{Shk09}.}
\tablefoottext{f}{New single young star candidate}.
}